\documentclass[acmsmall]{acmart}

\AtBeginDocument{%
  }

\usepackage{xcolor}
\usepackage{array}
\usepackage{booktabs}   
\usepackage{amsmath}    
\usepackage[htt]{hyphenat}
\usepackage{tabularx}
\usepackage{graphicx}
\usepackage{comment}
\usepackage{subfig}
\usepackage{subcaption} 
\usepackage{listings}
\usepackage{multirow}
\usepackage{fancybox}
\usepackage{hyperref}
\usepackage{algorithmic}
\usepackage{tcolorbox}
\usepackage{textcomp}
\usepackage{float}
\usepackage{mdframed}
\usepackage{soul} 
\usepackage{makecell}
\usepackage[normalem]{ulem}
\graphicspath{{figures/}}

\setcopyright{acmlicensed}
\copyrightyear{2027}
\acmYear{2027}
\acmDOI{XXXXXXX.XXXXXXX}
\acmConference[CSCW '27]{ACM Conference on Computer-Supported Cooperative Work
  and Social Computing}{2027}{Location TBD}
\acmISBN{978-1-4503-XXXX-X/2027/XX}

\begin{document}

\title{Longitudinal Multi-Dimensional Analysis of India-Pakistan Conflict Discourse on Reddit (2019-2025)}

\author{Yash Khare}
\email{khare.y@northeastern.edu}
\affiliation{%
  \institution{Northeastern University}
  \city{Boston}
  \state{MA}
  \country{USA}
}

\author{Hunjun Shin}
\email{shin.hu@northeastern.edu}
\affiliation{%
  \institution{Northeastern University}
  \city{Boston}
  \state{MA}
  \country{USA}
}

\author{Mohit Singhal}
\email{m.singhal@northeastern.edu}
\affiliation{%
  \institution{Northeastern University}
  \city{Boston}
  \state{MA}
  \country{USA}
}

\begin{abstract}
Six India--Pakistan crises between 2019 and 2025, from the Pulwama attack to Operation Sindoor, have each been argued out on Reddit in India-aligned communities such as \texttt{r/india}, in Pakistan-aligned communities such as \texttt{r/pakistan}, and in general-audience ones such as \texttt{r/worldnews}. A strike can be a security issue in one community and a question of who was harmed in another. Computational studies of the conflict have concentrated on the February 2019 crisis, with some splitting posts by national side, but to the best of our knowledge none has compared how differently aligned communities frame a series of crises. To investigate this, we collected 43K posts from ten such communities on Reddit and, after filtering, used a mixed-methods approach to analyze the 305K comments under the relevant posts. We label each comment's narrative frame with GPT-5.2 and measure contestation with Reddit's flag for comments that draw many votes in both directions. Contestation rates differ 2.5-fold between general-audience and India-aligned communities and 1.8-fold between humanitarian and economic framing, but only 1.1-fold between comments that Perspective scores as toxic and the rest. The mix of frames and sentiment tracks the kind of crisis more closely than its year: the first and last crises, both military and six years apart, draw nearly identical mixes, and the 2019 revocation of Article 370 draws a different one. Crossing into another group of communities sharpens opposition rather than softening it: commenters who post both inside and outside their usual group oppose the other side more often outside it, even as slightly fewer of their comments attack a person or group.
\end{abstract}



\maketitle

\providecommand{\candidate}[1]{}

\section{Introduction}
\label{sec:intro}

On 14 February 2019, a car bomb struck a paramilitary convoy at Pulwama, in Kashmir, killing at least 40 Indian personnel~\cite{aljazeera2019pulwama}. Twelve days later, the Indian Air Force (IAF) struck near Balakot, inside Pakistan~\cite{aljazeera2019balakot}. On 22 April 2025, gunmen killed 26 civilians near Pahalgam~\cite{aljazeera2025pahalgam}, and just over two weeks later India's military launched Operation Sindoor~\cite{aljazeera2025sindoor}. Pakistan answered with Operation Bunyan Marsoos~\cite{aljazeera2025bunyan}, and four days of missile, drone, and artillery exchanges between two nuclear-armed states ended in a ceasefire on 10 May~\cite{aljazeera2025ceasefire}. Each of these crises was argued out in public, at the same moment, in online communities belonging to neither government. On Reddit, \texttt{r/india}\footnote{\url{https://www.reddit.com/r/india/}} and \texttt{r/pakistan}\footnote{\url{https://www.reddit.com/r/pakistan/}} are the largest general-purpose communities for the two countries, each with its own moderators, its own rules, and its own voting public. Alongside them is \texttt{r/worldnews}\footnote{\url{https://www.reddit.com/r/worldnews/}}, one of the platform's largest international news communities, aligned with neither country. All three discuss a crisis between India and Pakistan at once.

A cross-border strike is one event, and the communities that discuss it produce more than one narrative~\cite{entman1991kal,tyagi2020ipconflict}. In one community, it arrives as a question of security and necessity; meanwhile, in another, as a question of who was harmed~\cite{makhortykh2017visual}. As the strikes began on 6 May 2025, a comment in \texttt{r/pakistan} reached a score of 153 by asking readers to ``look after each other and your own families\ldots some of us have never seen war in our lifetimes.'' Two days later, with the exchanges under way, a comment in \texttt{r/india} drew 182 for putting the question to its own government: ``\ldots Why were the villagers not evacuated then? Why allow the loss of life?''On 12 May, two days after the ceasefire, a comment in \texttt{r/worldnews} scored 1,151 with ``No more civilian casualties. There can be no greater victory than this.'' All three carry the same narrative frame, because each makes civilian harm the thing at stake. What each does with it differs, and so does the community it was posted in. We therefore treat community alignment as a property of the subreddit and not of any speaker's nationality, and the second comment above showed why: an India-aligned forum turning on its own government. This setting differs in one respect from the hostility within a single society that most harm research is built on~\cite{zampieri2019olid,davidson2017hatespeech,salehabadi2022user,aleksandric2025power}: alignment here follows a territorial dispute rather than a membership people choose~\cite{fang2022indivisibility}.

CSCW and HCI research has examined conflict discourse closely, but rarely by comparing the communities that argue over the same event. \citet{stewart2017lines} traced frame contests between two clusters of accounts, inferred from shared followers, in the discourse around one protest movement. \citet{zhou2025harm} annotated misinformation, narrative, and target in anti-Asian speech. \citet{olteanu2015crisis} compared separate crises rather than the communities arguing about a single one.  Our central construct is the frame, which \citet{entman1993framing} defines as selecting ``some aspects of a perceived reality'' and making them ``more salient.'' A narrative frame, in our usage, is that selection: the explanation a comment advances for the conflict, not the subject it mentions. A comment about an airstrike that turns on who was killed is humanitarian, while the same strike discussed as capability and deterrence is security and military. We use six such frames: political, security and military, religious and cultural, historical grievance, humanitarian, and economic.

Computational work on India--Pakistan discourse has concentrated on the weeks after a single crisis, the February 2019 Pulwama attack and the Indian airstrikes that followed it. \citet{tyagi2020ipconflict} propagated polarity labels through co-occurring hashtags to find polarizing tweets by Indian and Pakistani politicians, \citet{hussain2021pulwama} manually analyzed tweets under each country's leading hashtags, and \citet{palakodety2020hope} detected hope speech in YouTube comments. Each of these studies covers one crisis with one or two labels. Two of them divide tweets by national side, but none follows differently aligned communities across a series of crises. The closest Reddit study, of the Israeli--Palestinian conflict, labels stance alone, for about ten thousand comments, a thirtieth as many as we label~\cite{ali2025stance}.

To study a conflict that keeps returning, we follow six of its crises over six years. Four are military events in two attack-and-retaliation pairs: the Pulwama attack and the Balakot strike in February 2019, and the Pahalgam attack and Operation Sindoor in April and May 2025. The other two are the revocation of Article 370 in August 2019 and the ceasefire agreement of February 2021. Because military crises recur six years apart and 2019 includes two kinds of crises, we can separate the crisis type from the year it occurred. We retrieved submissions matching our crisis queries, along with their comments, from the Arctic Shift archive~\cite{arcticshift2026software} via the BAScraper client~\cite{bascraper2026software}, and kept submissions that gpt-5.2~\cite{openai2025gpt52} judged relevant to relations between the two countries. After filtering, we analyze 305,217 comments under submissions posted between January 2019 and June 2025 to ten communities. Six are India-aligned (\texttt{r/india}, \texttt{r/indianews}, \texttt{r/IndiaSpeaks}, \texttt{r/bakchodi}, \texttt{r/IndianDefense}, and \texttt{r/indiadiscussion}), one is Pakistan-aligned (\texttt{r/pakistan}), and three are general-audience communities in which this conflict is argued at scale (\texttt{r/worldnews}, \texttt{r/geopolitics}, and \texttt{r/lesscredibledefence}).

We describe each comment with six annotation layers. We chose gpt-5.2 after comparing commercial and open-weight models against human-coded references (Table~\ref{tab:benchmark}), and we use it to label content type, narrative frame, sentiment, and stance toward India and toward Pakistan. We label a comment as toxic if Google's Perspective API~\cite{jigsaw2018perspective} returns a TOXICITY score of 0.5 or higher, and we use a RoBERTa-large classifier~\cite{liu2019roberta} that we fine-tuned to decide whether a toxic comment is aimed at a target. We measure contestation with Reddit's \texttt{controversiality} flag~\cite{reddit2017source}, which the platform sets on comments that draw many votes in both directions. To the best of our knowledge, this is the first longitudinal study of India--Pakistan crisis discourse to span more than one crisis, and the first to model the narrative frame a comment advances together with the community that received it. We therefore treat a comment as a pairing of the frame it carries with the community it was posted to, and ask what that pairing explains.

In particular, we investigate the following research questions:

\begin{itemize}
\item \textbf{RQ1:} Which combinations of narrative frame and community make crisis discourse contested, and how do those combinations compare with the content-layer toxicity signal that harm research typically measures?
\item \textbf{RQ2:} Why do those combinations differ, and is an event's frame--affect signature fixed by the kind of crisis or by the year it occurred?
\item \textbf{RQ3:} What happens at the community boundary, where participants cross between opposed communities?
\end{itemize}

Our results show that what a community argues over depends on the frame a comment advances and the community it appears in, and hardly on how offensively it is worded. Humanitarian framing is the most contested category at 10.4\% and economic framing the least at 5.9\%; across the three alignment groups the range is wider, from 11.7\% in the general-audience communities to 4.7\% in the India-aligned ones. Community alignment separates controversiality rates 2.5-fold and narrative frame 1.8-fold, against 1.11-fold for the toxic-versus-not contrast that content moderation acts on. Measured against a toxicity threshold, toxic and non-toxic comments differ by 0.8 points, and inside humanitarian framing they barely differ at all, 10.2\% against 10.5\%; in a subreddit fixed-effects logit the frame gradient survives absorbing community composition while the toxicity coefficient stays negligible (odds ratio $1.17$, $p = 3.4\times10^{-11}$). 
The same frame is not the same object across communities. Under security and military framing, the adjusted probability that a comment reads as positive is 9.4\% in India-aligned communities against 5.0\% in general-audience ones, and the frame-by-community interaction \emph{strengthens} when stance toward both states is controlled ($\chi^2(20) = 287.6$, $p = 2.75\times10^{-49}$), so the asymmetry is not partisanship. A negatively read security and military comment receives a lower predicted score in \texttt{r/pakistan} than in India-aligned communities (1.239 against 1.418, on a sign-preserving log of Reddit's net vote count), with or without Operation Sindoor. The higher score in general-audience communities depends on that single event. An event's signature, meaning the mix of frames used and the sentiment attached to them, tracks the kind of crisis more closely than the year. \textit{Pulwama} and \textit{Operation Sindoor} are six years apart and have a cosine distance of 0.0074 (95\% CI [0.0022, 0.0313]), while Pulwama and the constitutional change six months later are at 0.2058. At the community boundary, crossing intensifies opposition instead of softening it. We found no discrete bridge-users; those who cross concentrate among the more hawkish participants ($\chi^2(2) = 299.5$, Cram\'er's $V = 0.226$, $p = 9.42\times10^{-66}$), and tracking 1,560 authors against themselves shows the same people raising their anti-out-group stance from 0.342 at home to 0.380 across the line ($+0.038$, $p < .001$, paired Wilcoxon).

In each of these results, the community is doing work that a comment-level measure cannot see. We make three contributions to CSCW, HCI, and social computing research on conflict discourse and online communities: 
\begin{itemize}
\item (1) \textit{Contestation tracks frame and community rather than tone} (RQ1). We show that what a community contests is a property of the frame and the community together, and hardly at all of a comment's tone. Humanitarian framing splits a community even when politely worded. Security and military framing draws it together. How offensive a comment is barely predicts whether a community will fight about it.
\item (2) \textit{A frame is not one object across communities} (RQ2). We find that the same frame is produced and rewarded differently depending on where it appears, and that an event's signature tracks the kind of crisis rather than the year. A frame label read without the community it appeared in loses what separates one community from another. 
\item (3) \textit{Crossing a community boundary is incursion rather than contact} (RQ3). No bridge-user type appears in the data. Those who cross skew hawkish, and the receiving community penalizes civil and hostile outsiders alike. Because we follow the same authors across the boundary, this is a change in how people write and not in who shows up. 
\end{itemize}
These findings bear on how harm is measured in conflict discourse, and on any design that routes people across community lines. For CSCW, our results carry a practical consequence: conflict discourse has to be measured at the pairing of a comment with the community that received it, because neither term predicts contestation on its own.

\section{Background and Related Work}

When India and Pakistan clash, communities aligned with each side and general-audience communities argue over the same event at once, and each makes something different of it. Three bodies of work bear on this: studies of how aligned audiences frame a shared conflict, of how harm and contestation in online discussion are measured, and of what happens when participants cross between opposed communities.

\subsection{Framing a shared conflict across aligned audiences}
\label{sec:rw-conflict}

\subsubsection{Frames and their inventories.}

Entman defined a frame by what it does: it selects ``some aspects of a perceived reality'' and makes them ``more salient in a communicating text, in such a way as to promote a particular problem definition, causal interpretation, moral evaluation, and/or treatment recommendation''~\cite{entman1993framing}. \citet{entman1991kal} showed that alignment shapes framing by comparing United States coverage of two civilian airliner shootdowns: one was covered as moral outrage and the other as a technical accident, a difference that tracked which side had fired. Frame effects depend on the relative strength of competing frames and on what an audience already believes~\cite{chongdruckman2007framing}, which gives a theoretical reason to expect one frame to behave differently in two communities. Computational work has operationalized frames as an inventory of emphases~\cite{card2015mediaframes,mendelsohn2021framing}.

\subsubsection{The same war, told by differently aligned audiences}

Differently aligned voices narrating the same event or issue emphasize different aspects, whether in online communities~\cite{makhortykh2017visual}, news ecosystems~\cite{hanley2023specialoperation}, or the commenters under publishers with different political leanings~\cite{rho2018metoo}. In the 2014 fighting in eastern Ukraine, for example, pro-Ukrainian communities framed the conflict as a limited operation against insurgents, and pro-Russian communities framed it as a war on the region's Russian population~\cite{makhortykh2017visual}. \citet{stewart2017lines} showed that such frame contests are networked, so the division can be read from network structure as well as from text. Frames also differ in how far they travel and how long they last, and they need not move with affect. Ukrainian memes that cast Ukrainians as victims drew roughly twice as many retweets as memes with no narrative~\cite{mejova2025narratives}, and on Reddit, framing of North Korea only partly reverted after the 2019 Hanoi summit failed, while sentiment fully reverted~\cite {shin2026diplomacy}. Studies that record alignment attach it to users or use it to group results. Twitter studies of framing on abortion~\cite{rao2025abortion}, a single campaign~\cite{panda2026framing} and the 2014 war in Ukraine~\cite{muradahmed2026ukraine} did one or the other, and Reddit studies of the Israeli--Palestinian conflict compared aligned and general-audience subreddits on a single attribute, such as stance or the extremity of opinion~\cite{ali2025stance,guerra2024extreme}. Researchers have studied discourse under sustained conflict through a drug war~\cite{dechoudhury2014narco}, the Iraq war as Iraqi bloggers recorded it~\cite{mark2012wardiary,alani2010blogging}, protests that turned into war~\cite{greger2025protests}, an invasion~\cite{corso2024longitudinal}, and repeated escalations within one conflict~\cite{zafeiropoulos2026telegram}. Most of these studies followed one conflict through one continuous window, so the period in which a pattern was recorded is hard to separate from the kind of event that produced it.

\subsubsection{The India--Pakistan conflict}
Framing research on this conflict predates computational work: among four Asian conflicts, Indian and Pakistani newspaper coverage of Kashmir showed the strongest war-journalism framing~\cite{lee2005war}. Computational work has focused on the 2019 crisis and has divided over what to look for. \citet{tyagi2020ipconflict} propagated polarity labels across a network of co-occurring hashtags after the Pulwama attack and found that politicians of India's governing party called for escalation more often than politicians of other parties, and a manual analysis of tweets on the same crisis divided them by national side~\cite{hussain2021pulwama}. \citet{palakodety2020hope} reversed the target and searched YouTube comments on the same events for hope speech. Work on code-mixed comments extended that search into Romanized Hindi~\cite{khudabukhsh2020codeswitch}, and the same approach later found sympathy in Pakistani tweets during India's second COVID-19 wave~\cite{yoo2021empathy}. The same events thus sustain both hostility and sympathy.

\begin{table*}[t]
\footnotesize
\caption{Computational work on this dyad and on the nearest adjacent cases.}
\label{tab:positioning}
\newcolumntype{Y}[1]{>{\hsize=#1\hsize\raggedright\arraybackslash}X}
\begin{tabularx}{\linewidth}{@{}Y{1.15}Y{0.70}Y{1.05}Y{1.30}Y{1.05}Y{1.03}Y{0.72}@{}}
\toprule
\textbf{Study} & \textbf{Platform} & \textbf{Events} & \textbf{Labels and method} & \textbf{Communities} & \textbf{Community in the measurement} & \textbf{Language} \\
\midrule
Tyagi et al.~\cite{tyagi2020ipconflict} & Twitter & Pulwama, Balakot & Hashtag labels, co-occurrence propagation & Politicians & Not modelled & English, Hindi, Urdu \\
Hussain et al.~\cite{hussain2021pulwama} & Twitter & Pulwama & Manual coding & Two national sides & Not modelled & English \\
KhudaBukhsh et al.~\cite{khudabukhsh2020codeswitch} & YouTube & Pulwama, Balakot & Code-switching, classifier & Not modelled & Not modelled & Code-mixed \\
Palakodety et al.~\cite{palakodety2020hope} & YouTube & Pulwama, Balakot & Hope speech, classifier & Not modelled & Not modelled & English \\
Yoo et al.~\cite{yoo2021empathy} & Twitter & India COVID-19 wave & Supportive speech; classifier & Country of origin (India, Pakistan, other) & Grouping variable & English \\
Ali et al.~\cite{ali2025stance} & Reddit & Israel--Palestine, 2023--24 & Stance (three classes), one layer & 12 subreddits & Grouping variable & English \\
Chen et al.~\cite{chen2024isamasred} & Reddit & Israel--Hamas, one window & Controversy flag, moral language & 100 subreddits & Grouping variable & English \\
Guerra et al.~\cite{guerra2024extreme} & Reddit & Israeli--Palestinian, 2023 & Extremity score & Four, incl.\ general-audience & Grouping variable & English \\
Rao et al.~\cite{rao2025abortion} & Twitter & Abortion, one year & Five frames, hostility; transformers & Liberal, conservative (per user) & Speaker attribute & English \\
Panda et al.~\cite{panda2026framing} & Twitter & One campaign & Grounded-theory frames; RoBERTa & One aligned public & Grouping variable & English \\
Murad Ahmed et al.~\cite{muradahmed2026ukraine} & Twitter & Ukraine 2014 & Themes, hashtag projection & Three blocs incl.\ neutral & Grouping variable & English \\
Zhou et al.~\cite{zhou2025harm} & Twitter & COVID-19 period & Misinformation, narrative, target & Anti-Asian hate & Not modelled & English \\
\midrule
\textbf{Our study} & \textbf{Reddit} & \textbf{Six, 2019--2025} & \textbf{Six annotation layers: gpt-5.2 against a codebook, Perspective, fine-tuned RoBERTa; human-validated} & \textbf{Ten, in three alignment groups} & \textbf{Measured: composed with frame, and the boundary crossed} & \textbf{English; Hindi or Urdu in 3.5\%} \\
\bottomrule
\end{tabularx}
\end{table*}

Prior studies of this conflict labeled the speaker on one or two attributes within a single crisis or period, and across them and the nearest adjacent cases, the community entered the design as a grouping variable at most. Table~\ref{tab:positioning} summarizes this. A grouping variable splits results by community, whereas we measure frame and affect within each community and model a comment's reception as a function of its frame and its community together, across ten communities in three alignment groups and six crises. RQ2 asks why those combinations of frame and community differ, and whether an event's frame--affect signature is set by the kind of crisis or by the year it occurred. The India--Pakistan conflict suits this design. We assign each community's alignment from the country it is organized around, so the alignment belongs to the community and not to the people who post there, and participants are audiences to one shared event on a clock set outside the platform, so comparing communities compares readings of one occurrence. Four of the six crises are military and form two attack-and-retaliation pairs six years apart, the 2019 pair of Pulwama and Balakot and the 2025 pair of Pahalgam and Operation Sindoor, so we can compare events of the same kind across years, which a single continuous window does not allow. To the best of our knowledge, no earlier study of this conflict has measured frame-level discourse with the receiving community as part of the measurement.

\subsection{Contestation, harm, and the community as a measured object}
\label{sec:rw-contestation}

\subsubsection{Composing harm.}

Harm is usually measured on the comment alone, as a single construct or a short hierarchy of them~\cite{zampieri2019olid,elsherief2021latenthatred,hartmann2025lostmoderation}. However, recent work argues that this is the wrong unit and toxicity is a contextual property of a comment within its conversation and community, which a text-intrinsic score discards by construction~\cite{berezin2025contextualharm,hartmann2026byebye}. While CSCW research has begun to compose such attributes, these compositions still overlook the community context. For instance, \citet{scheuerman2021severity} proposed a severity framework where dimensions of harm compound. Similarly, while \citet{zhou2025harm} composed misinformation, narrative, and target in anti-Asian hate speech, their analysis related harmfulness to the person judging rather than the venue. Neither composition includes the community in which a comment was posted.

\subsubsection{Norms at three levels.}

Reddit's norms operate at three levels. Moderator removals across a hundred large subreddits revealed macro norms enforced across most of the site, micro norms specific to a single subreddit, and between them meso norms shared by a group of subreddits but not by the site as a whole~\cite{chandrasekharan2018hiddenrules,singhal2023sok}. A comment acceptable in one subreddit can therefore be removed in another. These norms belong to the community: newcomers to political subreddits with distinctive toxicity levels nearly matched those levels in their first comments~\cite{rajadesingan2020norms}, and members described, in their own words, what their communities value in their own terms~\cite{weld2024values}. Crossmod, a moderation tool built on the three-level finding, implemented only the outer two, with one hundred classifiers for single subreddits, eight for site-wide norms, and none for a group of subreddits~\cite{chandrasekharan2019crossmod}. Recent work continues to find that toxicity and incivility signals alone do not capture these community-specific norms, and that even instruction-following LLMs struggle to enforce a community's own rules beyond universal violations~\cite{xin2024communityrules,kim2025rulebook}. 

\subsubsection{Reception, not hostility.}

Votes on Reddit signal reception for a particular post, but they are not a representative poll of a community's members. They come from whoever reads a comment, including non-members; votes themselves are scarce~\cite{gilbert2013underprovision}, and mobilized outsiders can arrive in numbers~\cite{kumar2018conflict}. A vote split therefore describes the reception a comment met in a community, which may differ from what its members think. Harm research, and the content moderation built on it, measures something else: how hostile a comment is~\cite{lees2022perspective}. The two need not coincide: nearly everyone can reject a hostile comment, and a civil one can split them, so a toxicity score does not indicate whether a community is divided over a comment.

\subsubsection{Measuring controversy.}

Controversy is often measured by how a discussion network splits~\cite{garimella2016controversy,garimella2018controversy} or, on Reddit, by a post's upvote ratio~\cite {hessel2019controversy}. For comments, Reddit sets its own \texttt{controversiality} flag once a comment's votes pass a minimum count and split roughly evenly~\cite{reddit2017source}. Reddit does not publish the values it uses, so, like the reporting flags platforms offer users~\cite{crawford2016flag}, the flag is as much a platform artifact as a record of contestation. \citet{chen2024isamasred} released the flag with roughly eight million comments on the Israel--Hamas conflict and reported flag rates by subreddit; a study of Reddit discussion cascades also found that those starting from flagged content spread faster and further~\cite{jasser2021controversial}. However, neither study related the flag to a comment's frame.

\subsubsection{Frame, community and contestation.}  
Studies that related a frame to its reception measured engagement, such as retweets~\cite{mendelsohn2021framing,mejova2025narratives}, not divided reception within a community whose alignment was fixed in advance, and the harm compositions above leave the community out. We compose frame with community and take contestation, inferred from the flag, as the outcome. Hostility is a second outcome, compared with contestation rather than standing in for it. For the community term, we hypothesize that the communities in each of our alignment groups (India-aligned, Pakistan-aligned and general-audience) form a meso-level group, sharing norms about what they reward that neither the site as a whole nor any single subreddit captures.

\subsection{Crossing the community boundary}
\label{sec:rw-crossing}

\subsubsection{Friction at the boundary.}
Contact theory expects contact between groups to reduce prejudice, especially when groups meet as equals, share goals, cooperate, and have institutional support~\cite{pettigrew2006contact}. Bad contact weighs more than good contact: negative contact makes group membership more salient~\cite{paolini2010negative} and predicts prejudice more strongly than positive contact predicts its reduction~\cite{barlow2012contact}. Posting into a community aligned with the other side of a shooting war is contact under none of the favorable conditions. Where online contact between communities has been measured, most of it was not hostile, but the hostile share was concentrated and costly. Fewer than one percent of Reddit communities started three-quarters of the conflicts identified as negative mobilizations, and those conflicts reduced activity in the targeted communities~\cite{kumar2018conflict}. Across ideological lines, exchanges on r/politics were more negative than those between like-minded users, and more likely to be cut short when one user spoke negatively of the other's side~\cite{marchal2022benice}. Exposure experiments found similar limits: following a bot that retweeted the other party's elites made Republicans more conservative~\cite{bail2018exposure}, and exposure to a shared external adversary during the 2020 US--Iran confrontation increased polarisation among Republican participants~\cite{jahani2022commonenemy}.

\subsubsection{Where contact goes differently.}
Contact does not always go this way, and its outcome depends on conditions~\cite{xia2025integrated}. Opposed partisans on Reddit did not avoid each other, and toxicity did not concentrate across party lines: supporters of the two 2016 US presidential candidates replied to each other more often than a random baseline predicted~\cite{defrancisci2021noecho}, and toxic speech was more common between political communities on the same side than between opposed ones, though mainly as shared hostility toward outgroups~\cite{efstratiou2023nonpolar}. Design can also help. Reddit users described strategies and designs that could help sustain cross-partisan discussion~\cite{rajadesingan2021crosspartisan}, and cross-party conversations sharply reduced affective polarisation when they avoided disagreement, although the effect decayed over time~\cite{santoro2022crosspartisan}. How a crosser writes matters as well as which side they come from: cross-cutting replies that expressed warmth toward the other side tended to draw positive responses~\cite{marchal2022benice}, and antisocial behavior is partly situational rather than dispositional~\cite{cheng2017troll}. The spiral of silence offers a rival account of who crosses at all: in a hostile opinion climate, only a committed hard core keeps speaking~\cite{noelleneumann1974spiral}.

\subsubsection{Measuring a crosser.}
Measured studies of crossers have mostly followed the crosser's own hostility. Cross-partisan replies on YouTube were more toxic than co-partisan ones, and conservatives crossed far more often than liberals~\cite{wu2021crosspartisan}, while most users who comment in several communities changed their toxicity across communities~\cite {almerekhi2022toxicity}. These designs measured what crossers wrote, not how the receiving community responded, so they could not show whether a community penalizes outsiders. \citet{datta2019interconflict} measured responses by inferring conflict between subreddits from authors who were upvoted in one subreddit and downvoted in another. They defined each author's home by where the author was rewarded and inferred opposition between communities rather than fixing it in advance. Crossers are also usually treated as a kind of person rather than a matter of degree, although activity across communities is continuous, with generalists and specialists at opposite ends of a spectrum~\cite{waller2019generalists}. 

\subsubsection{What this means for design.}
For CSCW, this is a design question about what platform designers or a community's own moderators could change~\cite{rajadesingan2021crosspartisan}, and its answer depends on two things these studies leave open. The first is what a receiving community penalizes: an outsider's side, their register (the tone and vocabulary a comment is written in) or simply being an outsider. The second is whether crossing changes how a person writes or only who crosses. We address the first by comparing how opposed communities score outsiders who differ in side and register, and the second by comparing each author's comments in their own community with those the same author posts across the boundary. We describe crossers by regions of the continuum between generalists and specialists rather than by a threshold, and our finding that the authors who cross come disproportionately from the more combative end of the corpus is consistent with the spiral-of-silence expectation~\cite{noelleneumann1974spiral}.

\section{Data and Methods}

\label{sec:data}

\subsection{Corpus}
\label{sec:data-site}
\label{sec:data-site-reddit}

Researchers have used Reddit to study political polarization~\cite{waller2021quantifying}, echo chambers~\cite{cinelli2021echo}, the rules communities enforce~\cite{chandrasekharan2018hiddenrules}, and the framing of international affairs~\cite{shin2026diplomacy}. Each subreddit has its own moderators and rules, so the boundaries between the communities we compare are recorded by the platform itself. Each comment is threaded and carries its author's name, so the same author can be followed from one community to another, as earlier work measured conflict between subreddits~\cite{kumar2018conflict,datta2019interconflict}. We therefore chose Reddit for this study. 

\begin{figure}[!h]
  \centering
  \includegraphics[width=\textwidth,alt={The pipline for our corpus generation.}]{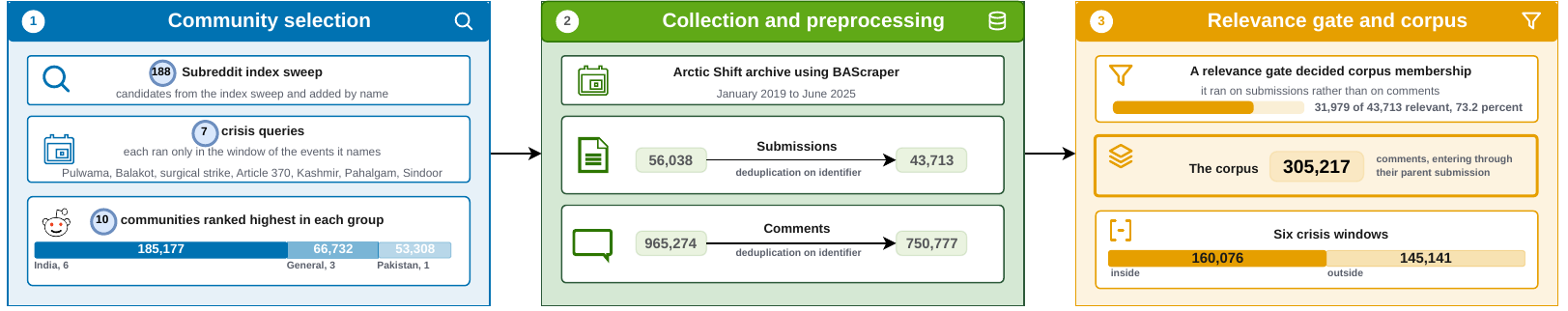}
  \caption{How we built the corpus. Left: the archive check of the communities. Center: retrieval and deduplication of posts and comments. Right: the relevance gate, the resulting corpus and its split across the six crisis windows.}
  \label{fig:pipeline}
\end{figure}

Figure~\ref{fig:pipeline} summarizes how we built the corpus. We chose the communities in two steps: we assembled candidates, then ranked them by how much they discuss the crises. To assemble candidates, we searched the subreddit index of the Arctic Shift archive~\cite{arcticshift2026software} for community names beginning with regional stems (e.g., \texttt{india}, \texttt{pakistan}, \texttt{kashmir}, \texttt{desi}) or with stems for international news, geopolitics and defense (e.g., \texttt{worldnews}, \texttt{geopolitic}, \texttt{defence}, \texttt{defense}), and kept those with at least 5,000 subscribers, since a smaller community is unlikely to hold much discussion of the crises. The index matches only the start of a name, so the search cannot find a relevant community whose name begins differently. We therefore added these communities by name from our knowledge of the platform, including \texttt{r/bakchodi}, a large India-focused community, and \texttt{r/lesscredibledefence}, a community for defense analysis. Studies of online communities often combine an automated search with communities the researchers know to be relevant~\cite{ribeiro2020auditing,hortaribeiro2021manosphere}, a form of purposive sampling~\cite{palinkas2015purposeful}. Because the search found far fewer Pakistan-focused communities than India-focused ones, we also added six Pakistan-focused communities that we had examined by hand (\texttt{r/karachi}, \texttt{r/Lahore}, \texttt{r/pakistani}, \texttt{r/pkmigrate}, \texttt{r/PakistaniMilitary} and \texttt{r/PakistanDiscussions}); the last two had fewer than 5,000 subscribers, to search the Pakistan side as widely as we could. Even so, the resulting pool of 188 candidates holds 20 Pakistan-focused communities against 147 India-focused ones. 

To rank them, we ran seven crisis queries against every candidate, each restricted to the window of its event: ``\textit{Pulwama}'', ``\textit{Balakot}'' and ``\textit{surgical strike}'' in February and March 2019; ``\textit{Article 370}'' and ``\textit{Kashmir}'' in August 2019; and ``\textit{Pahalgam}'' and ``\textit{Operation Sindoor}'' in April and May 2025. Restricting each query to its window counts discussion of the crisis itself rather than every use of a common word such as ``\textit{Kashmir}''. We ranked the candidates by the number of posts these queries matched. Raw volume favors large communities. Table~\ref{tab:selection} lists each community's subscriber count beside its matches, separating small communities focused on the conflict from very large ones that rarely mention it.
\begin{table}[!h]
\centering
\caption{Candidate communities ranked by crisis-period volume, with the fourteen collected marked.}
\label{tab:selection}
\setlength{\tabcolsep}{4pt}
\resizebox{0.65\columnwidth}{!}{%
\begin{tabularx}{0.85\linewidth}{@{}>{\raggedright\arraybackslash}Xrrcc@{}}
\toprule
\textbf{Community} & \textbf{Subscribers} & \textbf{Query Matches} & \textbf{Data Collected} \\
\midrule
\multicolumn{4}{@{}l}{\itshape India-aligned} \\
\hline
\texttt{r/india} & 2,482,127 & 1,995 & \checkmark \\
\texttt{r/indianews} & 211,511 & 591 & \checkmark \\
\texttt{r/IndiaSpeaks} & 1,020,810 & 438 & \checkmark \\
\texttt{r/bakchodi} & 105,209 & 224 & \checkmark \\
\texttt{r/IndianDefense} & 24,893 & 204 & \checkmark \\
\texttt{r/indiadiscussion} & 102,231 & 194 & \checkmark \\
\hline
\addlinespace
\multicolumn{4}{@{}l}{\itshape General-audience} \\
\hline
\texttt{r/worldnews} & 44,166,945 & 1,421 & \checkmark \\
\texttt{r/geopolitics} & 687,192 & 128 & \checkmark \\
\texttt{r/lesscredibledefence} & 28,885 & 101 & \checkmark \\
\hline
\addlinespace
\multicolumn{4}{@{}l}{\itshape Pakistan-aligned} \\
\hline
\texttt{r/pakistan} & 423,985 & 833 & \checkmark \\
\texttt{r/karachi} & 96,480 & 11 & \\
\texttt{r/PakistanDiscussions} & 112 & 7 & \\
\texttt{r/IslamabadSocial} & 12,338 & 5 & \\
\texttt{r/PakistaniiConfessions} & 57,329 & 3 & \\
\texttt{r/islamabad} & 141,190 & 2 & \\
\texttt{r/PakistaniMilitary} & 3,361 & 1 & \\
\bottomrule
\end{tabularx}
}
\end{table}
We analyze ten communities, the highest-ranked in each alignment group. They fall into three groups, defined by the country each community is organized around: six India-aligned communities, the six highest-volume candidates on the Indian side, which hold 185,177 comments; three general-audience communities of international news, geopolitics and defense, which hold 66,732; and one Pakistan-aligned community, \texttt{r/pakistan}, which holds 53,308. Two authors assigned each community to a group independently, based on its name, description, and rules. They agreed on 161 of the 188 communities (Cohen's $\kappa = 0.73$), showing \textit{substantial agreement}, and resolved disagreements through discussion. The ordering is decisive on the Pakistan side: \texttt{r/pakistan} returned 833 matching posts, and none of the other nineteen Pakistan-focused candidates returned more than 11 (Table~\ref{tab:selection}).

Two of the seven queries, ``surgical strike'' and ``Operation Sindoor'', use India's names for its own operations, so we repeated the Pakistan-side check with the names Pakistan's government and press use for the same events, among them ``Operation Swift Retort'', ``Bunyan-un-Marsoos'', ``Marka-e-Haq'', and ``IIOJK'' for Indian-administered Kashmir. We matched both sets of names against the title and body of every post posted in the three windows to eight Pakistan-focused communities. Apart from \texttt{r/pakistan}, none reached 40 matching posts even when we combined both sets; the largest were \texttt{r/karachi} (35) and \texttt{r/islamabad} (22), and other candidates posted nothing during the 2019 windows. On Reddit, discussion of these crises in Pakistan-focused communities is concentrated in \texttt{r/pakistan}. We therefore treat \texttt{r/pakistan} as a single case, compare it with \texttt{r/india} and the spread across the six India-aligned communities, and never treat it as a group. We draw every alignment-group result in this paper from the 305,217 comments in these ten communities. An alignment describes a community organized around one country and says nothing about the nationality or views of the people who comment in it, so we draw no conclusions about an individual commenter from a community-level result.

\subsection{Data collection and preprocessing}

We collected posts and their comments from the Arctic Shift archive~\cite{arcticshift2026software} through the BAScraper client~\cite{bascraper2026software}. In each of the ten communities, we searched the titles and bodies of posts posted between January 2019 and June 2025 for any of 39 keyword queries, which name the two countries together, Kashmir, the attacks, military action, the border and the ceasefire, or nuclear weapons (Appendix~\ref{app:queries} lists the queries and the posts each matched). Our collection window opens one month before the first of our six crises (Pulwama, in February 2019) and closes a month after the last (Operation Sindoor, in May 2025), so that the first and last crises, like the four between them, have ordinary discussion on either side. We then downloaded every comment under each matched post; the most recent was posted on 25 January 2026. A post could match several queries, so we removed duplicate posts and comments by their Reddit identifiers, which left 43,713 posts and 750,777 comments (Table~\ref{tab:funnel}).

\begin{figure}[htb]
  \centering
  \includegraphics[width=.85\linewidth,alt={Annoted comments spread per month.}]{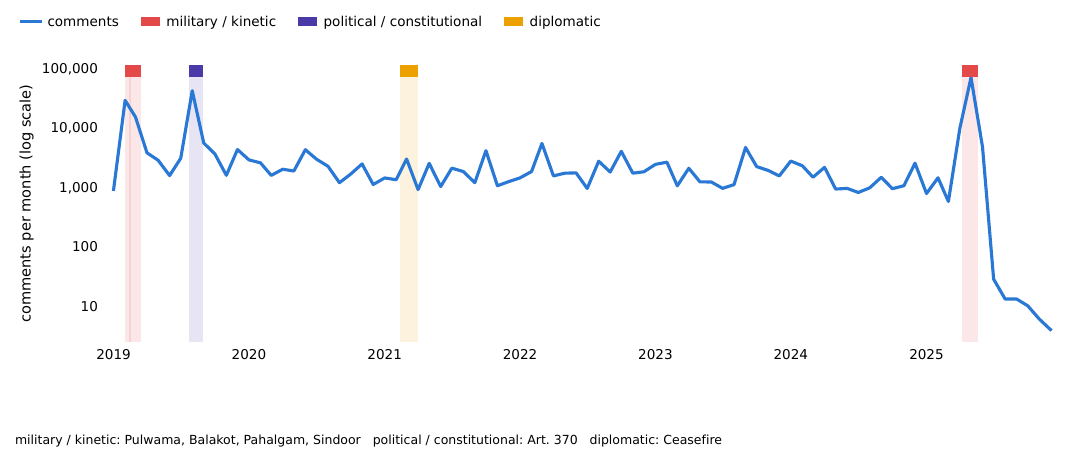}
  \caption{Annotated comments per month from January 2019, on a logarithmic axis. Shaded bands mark the six crisis windows, colored by crisis type: military or kinetic (Pulwama, Balakot, Pahalgam, and Operation Sindoor), political or constitutional (Article 370), and diplomatic (the 2021 ceasefire).}
  \label{fig:timeline}
\end{figure}

Broad queries also match posts that are not about relations between India and Pakistan: ``ceasefire'' and ``nuclear'', for instance, also match threads about other countries. Narrower queries would have cut these matches but would have also missed relevant posts that use the same words, and posts the queries miss never reach our corpus. We therefore kept the queries broad and screened every match with a relevance gate. In a zero-shot prompt that states our criteria for relevance, we ask gpt-5.2~\cite{openai2025gpt52} whether a post engages with relations between the two countries, and we give it the post's title, the first 100 words of its body, its subreddit, the crisis type and phase it falls in (Figure~\ref{fig:timeline}), and its score and comment count. The prompt lets the score and comment count settle a case only when the title and body leave it borderline. Appendix~\ref{app:codebook} states these criteria, and Appendix~\ref{app:prompts} summarizes the system and user prompts. We label every post through OpenAI's Batch API~\cite{openai2026batchapi}. The gate keeps 31,979 posts (73.2\%) and removes 11,734 (26.8\%). The share it keeps is lowest in the three general-audience communities (29\% in \texttt{r/lesscredibledefence}, 44\% in \texttt{r/geopolitics}, and 67\% in \texttt{r/worldnews}) and highest in \texttt{r/IndianDefense} (86\%). 

\begin{table}[htbp]
\centering
\caption{Posts and comments at each preprocessing step.}
\label{tab:funnel}
\footnotesize
\setlength{\tabcolsep}{3pt}
\begin{tabularx}{\linewidth}{@{}>{\raggedright\arraybackslash}Xrrr@{}}
\toprule
Step & Posts & Comments & Comments removed \\
\midrule
After deduplication & 43,713 & 750,777 & \\
Relevance gate & 31,979 & 432,495 & 318,282 (42.4\%) \\
Bot accounts and messages & & 427,546 & 4,949 (1.1\%) \\
Deleted or removed text & & 340,609 & 86,937 (20.1\%) \\
Empty body & & 331,485 & 9,124 (2.1\%) \\
Under three words & & 305,217 & 26,268 (6.1\%) \\
\midrule
Final corpus & 9,186 & 305,217 & 127,278 (29.4\%) \\
\bottomrule
\end{tabularx}
\end{table}

To check the gate, two coders independently labeled 200 posts stratified into five tiers of expected difficulty, from clear cases to ambiguous ones, which we assigned by how strongly each post matched our keyword queries. The coders agreed on 183 (91.5\%, Cohen's $\kappa = 0.818$), showing \textit{almost perfect agreement}. We scored the gate's production labels against these resolved labels. The gate agrees at $\kappa = 0.855$ (95\% CI 0.753 to 0.937); on the relevant class its precision is 0.971, its recall 0.904, and its $F_1$ 0.936. Seven of its nine errors turned away a relevant post, and two let an irrelevant one through, so its mistakes cost us coverage more than they add noise. The sample over-represents hard cases, so these figures describe the sample rather than the whole collection. It also holds only 11 posts from \texttt{r/pakistan}, two of which the gate wrongly excluded, which is too few for a separate estimate. In a benchmark on all 200 posts, gemini-3-pro-preview and claude-opus-4.6 agreed slightly more often with the reference than gpt-5.2 did ($\kappa = 0.895$ for both, against 0.874). The difference is two correct labels out of 200, too few for an exact McNemar test on the paired labels to detect ($p \geq 0.5$).

Many comments fit only within their thread, so we judged relevance once per post and let its comments inherit the label. The gate therefore removed 318,282 comments (42.4\%) and kept 432,495, including any off-topic replies under relevant posts. We then removed comments written by bots or left without usable text; each share below is of the 432,495 (Table~\ref{tab:funnel}). We identified bots from a list of 11 known accounts, such as AutoModerator and RemindMeBot, and from a regular expression that matches automated-message text; their 4,949 comments (1.1\%) are not participant discussion and would inflate activity counts. Another 86,937 (20.1\%) had lost their text: moderators or Reddit removed 72,148, and their authors deleted 14,789. Lost text is more common in \texttt{r/pakistan} (27.3\%) than in the India-aligned communities (16.6\%), and twice as common in 2025 as in 2019, so moderation and deletion have filtered the remaining comments unevenly. We kept all 8,148 comments whose authors later deleted their accounts, because their text survives and dropping them would remove people who later left Reddit. We also dropped 9,124 comments (2.1\%) with an empty body.

Before labeling, we replaced links, user, and community names with placeholders, so the text we sent to the language models contained no links or tagged user names. We also stripped Reddit formatting and quoted lines, because a quoted line repeats someone else's words and would otherwise be labeled as the commenter's own. We then dropped the 26,268 comments (6.1\%) with fewer than three words. Such comments usually hold too little text to code for frame or stance, although the rule also drops short slogans that state a stance outright. We did not filter by language. Standard language identifiers fail on code-mixed social media text~\cite{das2014identifying}, so we wrote a rule-based detector in the manner of dictionary-based language identification for such text~\cite{barman2014code}. It flags a comment that contains Devanagari or Perso-Arabic characters or at least two words from a list of 104 common romanized Hindi and Urdu words. Using this detector, 3.5\% of the comments in our final corpus contain Hindi or Urdu and 0.2\% use one of those two scripts; the other 96.5\% show no sign of either. A language filter would have removed more comments from the India-aligned communities, where 4.9\% contain Hindi or Urdu, than from \texttt{r/pakistan} (2.8\%) or the general-audience communities (0.2\%), and would have skewed the comparison between groups. These steps removed 127,278 comments (29.4\%), leaving us with the corpus we analyze: 305,217 comments under 9,186 posts. Of the 31,979 relevant posts, 21,888 have no comments in our collection. Reddit's own count is zero for 10,607 of them, and the other 11,281 list 55,495 comments that the archive did not return.

Our corpus is concentrated around the six crises. Figure~\ref{fig:timeline} places their time windows: comment volume rises by orders of magnitude inside them and falls back between them. Each window opens up to two weeks before its event. We call those days the baseline phase, the day of the event and the three days after it the acute phase, and the rest of the window, which runs to between two and five weeks after the event, the extended phase. Pulwama and Pahalgam have no extended phase, because the next event followed within about two weeks. Four of the events form two attack-and-retaliation pairs of the same type, Pulwama with Balakot in 2019 and Pahalgam with Operation Sindoor in 2025. The other two, the revocation of Article 370 in 2019 and the 2021 ceasefire, are each the only event of their type, so for those two types we cannot separate the type's effect from the event's. The windows differ by more than an order of magnitude, with 68,644 comments for Operation Sindoor versus 3,880 for the 2021 ceasefire. Of the 305,217 comments in our corpus, 160,076 fall inside a window and 145,141 outside. Most of our results use the 160,050 in-window comments that have both a sentiment and a frame label, which we call our analytic sample. Of these, 96,618 (60.4\%) are in India-aligned communities, 44,599 in general-audience communities, and 18,833 in \texttt{r/pakistan}. Because event volumes differ so much, we compare events by proportion within a community~\cite{olteanu2015crisis}, and we repeat every headline result without Operation Sindoor.

\subsection{Measures}

Our measure of contestation is Reddit's \texttt{controversiality} flag, which the platform sets on a comment that draws many votes in both directions. Reddit sets the flag once a comment's votes pass a minimum count and its share of upvotes falls within a band around one half; the example configuration in Reddit's source code, public until 2017, set these at seven votes and a share between 0.4 and 0.6~\cite{reddit2017source}. In our analytic sample, 7.1\% of comments carry it.

Beyond contestation, we describe each comment with six annotation layers (Table~\ref{tab:layers}). Four describe what it says: its content type, its narrative frame, its sentiment, and its stance toward India and toward Pakistan, coded as two separate axes. Two describe toxicity: whether the comment is toxic, and, if it is, whether the toxicity is aimed at a target. Our codebook, in Appendix~\ref{app:codebook}, follows the prescriptive paradigm~\cite{rottger2022paradigms}: each layer has an explicit decision rule and a tie-break, so our coders apply one reading at a contested boundary rather than reporting their own.

\begin{table}[tbp]
\centering
\caption{Annotation constructs, label distributions and mean model confidence.}
\label{tab:layers}
\begin{tabular}{llrrr}
\toprule
Construct & Label & N & Share & Mean confidence \\
\midrule
content type & Opinion/Commentary & 243337 & 78.7 & 0.878 \\
content type & Question/Query & 51063 & 16.5 & 0.877 \\
content type & Satire/Humor & 9202 & 3.0 & 0.871 \\
content type & Factual Claim & 5600 & 1.8 & 0.756 \\
sentiment & Negative & 193470 & 62.6 & 0.807 \\
sentiment & Neutral & 90909 & 29.4 & 0.732 \\
sentiment & Positive & 24781 & 8.0 & 0.757 \\
stance toward India & Neutral-India & 190554 & 61.6 & 0.593 \\
stance toward India & Pro-India & 61540 & 19.9 & 0.720 \\
stance toward India & Anti-India & 57109 & 18.5 & 0.735 \\
stance toward Pakistan & Neutral-Pakistan & 220552 & 71.3 & 0.676 \\
stance toward Pakistan & Anti-Pakistan & 70695 & 22.9 & 0.772 \\
stance toward Pakistan & Pro-Pakistan & 17956 & 5.8 & 0.719 \\
narrative frame & Political & 132353 & 42.8 & 0.629 \\
narrative frame & Security/Military & 112821 & 36.5 & 0.733 \\
narrative frame & Religious/Cultural & 20585 & 6.7 & 0.740 \\
narrative frame & Historical Grievance & 15141 & 4.9 & 0.700 \\
narrative frame & Humanitarian & 14332 & 4.6 & 0.704 \\
narrative frame & Economic & 13961 & 4.5 & 0.726 \\
offensiveness & Not Offensive & 270537 & 89.9 & 1.000 \\
offensiveness & Offensive-Targeted & 27473 & 9.1 & 0.993  \\
offensiveness & Offensive-Untargeted & 2909 & 1.0 & 0.947 \\
\bottomrule
\end{tabular}
\vspace{2pt}
\end{table}

Our content-type layer separates opinion from questions, satire, and factual claims. The gpt-5.2 labeler we deployed did not reproduce this four-way distinction. On the items both coders agreed on, it reaches only $\kappa = 0.432$ (Table~\ref{tab:reliability}) and recovers only 1 of the 14 factual claims in the adjudicated reference. We therefore use only its coarsest split, opinion against everything else. On the same agreed items, the labeler finds opinion reliably ($F_1 = 0.92$) but not the rest ($F_1 = 0.51$), and its $\kappa$ on this split (0.43) is no higher than on the four labels, so we never use content type as an outcome. We use it only as a control and stratifier, separating opinion, which is 79\% of the corpus, from the rest.

We measure toxicity with the TOXICITY attribute of Google's Perspective API~\cite{jigsaw2018perspective,lees2022perspective} and label a comment toxic if it scores 0.5 or higher, a threshold used in earlier work~\cite{rajadesingan2020norms,trujillo2022reddit,saveski2021structure,aleksandric2025power,kumarswamy2025causal}. To check the threshold on our data, two coders labeled 200 comments sampled from both sides of it and resolved their disagreements through discussion. The two coders agree at only $\kappa = 0.477$. Toxicity is a judgment on which annotators often disagree, partly because it depends on their beliefs and identities~\cite{sap2022annotators,davani2022dealing}. Perspective's 0.5 threshold agrees with the reviewer-resolved labels at $\kappa = 0.670$ (Table~\ref{tab:reliability}). These two figures answer different questions: the first measures how far two people agree, and the second how closely the cut matches the resolved labels. Above this boundary, we separate targeted from untargeted toxicity, following the offensive-language taxonomy of \citet{zampieri2019olid}: a toxic comment is targeted if it attacks an identifiable person, group, nation, or religion. We fine-tuned RoBERTa-large~\cite{liu2019roberta} on 500 toxic comments that our coders labeled for target type (Table~\ref{tab:reliability}) and used it to make this second decision for all 30,044 comments Perspective scored as toxic. In five-fold cross-validation on those comments, it agrees with the resolved labels at $\kappa = 0.487$ ($\pm 0.117$ across folds). The cross-validated agreement is moderate and well below the coders' 0.682 on the same 500 comments, so the classifier often places a toxic comment on the wrong side of the line between targeted and untargeted. A small difference in targeted toxicity between groups or periods may therefore reflect classifier error rather than a real difference.

\subsection{Human annotation}
\label{sec:data-coders}

Besides the relevance and toxicity checks, we built the human references in Table~\ref{tab:reliability} from three comment samples: 250 for content type with narrative frame, 250 for sentiment with both stance axes, and 500 for target type from comments that Perspective scores as toxic. We stratified the content-type, frame, sentiment, and stance samples by crisis type, alignment group, and crisis phase, oversampling the smaller alignment groups. For target type, we took two comments from each community, then sampled by quartile of Perspective's identity-attack and insult scores, balanced half and half between comments a Perspective-based rule marked as targeted and as untargeted. Stratifying over-represents rare categories and makes the samples harder to code than a random draw. Three coders took part, and two of them labeled every item independently in Label Studio~\cite{tkachenko2020labelstudio}. On each screen, we showed them one comment with the post it replied to (for relevance, the post alone), metadata that included its community and the crisis event, and one panel per layer with that layer's decision rule printed above the options.

We paired the three coders differently across projects. Our codebook asks for reading competence in English and in Hindi or Urdu, and for familiarity with South Asian politics. Two coders met both conditions and coded target type and the toxicity check together. The third read English and coded Hindi and Urdu text from Google translations; we paired this coder with one of the other two for relevance, content type, narrative frame, sentiment, and both stance axes. Text that needed translating was rare: in each content-type and sentiment sample, 7 of the items from our ten communities contain Hindi or Urdu, and leaving them out moves $\kappa$ by at most 0.023 on content type, frame, sentiment, or either stance axis. Most comments are in English but turn on local references, so the third coder's main gap is regional context rather than language.

\begin{table}[!h]
\centering
\caption{Agreement between our two coders, and between each layer's deployed
labeller and the coders' labels.}
\label{tab:reliability}
\small
\setlength{\tabcolsep}{3pt}
\begin{tabularx}{\linewidth}{@{}>{\raggedright\arraybackslash}Xrrrcl r@{}}
\toprule
 & \multicolumn{4}{c}{two coders} & \multicolumn{2}{c}{deployed labeller} \\
\cmidrule(lr){2-5}\cmidrule(l){6-7}
layer & $n$ & raw & $\kappa$ & 95\% CI & system & $\kappa$ \\
\midrule
relevance & 200 & 0.913 & 0.818 & [0.711, 0.914] & gpt-5.2 & 0.893 \\
content type & 250 & 0.872 & \textbf{0.619} & [0.471, 0.748] & gpt-5.2 & 0.432 \\
narrative frame & 250 & 0.797 & \textbf{0.700} & [0.615, 0.778] & gpt-5.2 & 0.775 \\
sentiment & 250 & 0.733 & 0.541 & [0.426, 0.643] & gpt-5.2 & 0.799 \\
stance toward India & 250 & 0.717 & 0.535 & [0.429, 0.632] & gpt-5.2 & 0.825  \\
stance toward Pakistan & 250 & 0.783 & 0.579 & [0.466, 0.685] & gpt-5.2 & 0.830 \\
target type & 500 & 0.893 & \textbf{0.671} & [0.589, 0.753] & RoBERTa & 0.487\textsuperscript{b} \\
toxicity at 0.5 & 200 & 0.745 & 0.477 & [0.360, 0.588] & Perspective & 0.670\textsuperscript{a} \\
\bottomrule
\end{tabularx}
\vspace{2pt}
{\footnotesize \textsuperscript{a}The 0.5 threshold against the reviewer-resolved labels of all
200 items. \textsuperscript{b}Five-fold cross-validation on all 500 sampled
items; the deployed classifier was then trained on all of them.\par}
\end{table}

Table~\ref{tab:reliability} reports two kinds of agreement for each layer. The first is between our two coders, on the items from our ten communities, with a 95\% confidence interval from 2,000 bootstrap resamples. Intercoder reliability ranges from $\kappa = 0.477$ for the toxicity check to 0.818 for relevance. On the scale of Landis and Koch~\cite{landis1977measurement}, relevance reaches almost perfect agreement (above 0.80), content type, narrative frame, and target type reach substantial agreement (0.61 to 0.80, in bold in the table), and sentiment, both stance axes, and the toxicity check reach moderate agreement (0.41 to 0.60). The second is between the coders' labels and the system whose labels we analyze: gpt-5.2 for relevance and the four comment layers it labels, our RoBERTa-large classifier for target type, and Perspective's 0.5 threshold for toxicity. For gpt-5.2, the count in parentheses is the number of items both coders agreed on and that we could match to our corpus.

\subsection{Choosing and deploying a labeler}
\label{sec:data-labeller}

Before adopting a labeler, we compared commercial and open-weight language models, encoders, and a lexicon baseline against the adjudicated references (Table~\ref{tab:benchmark}, details in Appendix~\ref{app:benchmark}). The table's coder values differ slightly from those in Table~\ref{tab:reliability} because its samples also include communities outside our ten. On sentiment and stance, gpt-5.2, gemini-3.1-pro-preview, and claude-opus-4.6 each agreed with the references more closely than any open-weight system did, although the three also voted on the disputed items when we built those references, which favors them in this comparison. We label content type, narrative frame, sentiment, and both stance axes for the whole corpus with gpt-5.2 through OpenAI's Batch API, because, of the GPT-5 models we tested, it scored highest on content type and frame, matched the other commercial models on sentiment and stance, and lets one model label the gate and all four of these layers. Averaged over sentiment and both stance axes, $\kappa$ was 0.720 for gpt-5.2, 0.725 for gemini-3.1-pro-preview, and 0.734 for claude-opus-4.6, no model led on every layer, and on each layer they differed by at most 0.10. We set reasoning effort to none and temperature to zero, use a strict output schema, and ask for both stance axes in one call. For each label, gpt-5.2 also reports a confidence, which our robustness analysis uses as one arm of a specification curve. 

\begin{table}[htbp]
\centering
\caption{Agreement of each labeler we tested with the adjudicated reference
for each layer, beside the agreement between our two coders.}
\label{tab:benchmark}
\small
\setlength{\tabcolsep}{3pt}
\begin{tabularx}{\linewidth}{@{}ll>{\raggedright\arraybackslash}Xrr@{}}
\toprule
layer & system & approach & $\kappa$ & $n$ \\
\midrule
content type & two coders &  & 0.612 & 250 \\
 & gpt-5.2 & zero-shot & 0.599 & 250 \\
\midrule
narrative frame & two coders &  & 0.665 & 250 \\
 & gpt-5.2 & zero-shot & 0.688 & 250 \\
\midrule
sentiment & two coders &  & 0.549 & 250 \\
 & VADER & lexicon & 0.157 & 250 \\
 & XLM-T & pre-trained & 0.374 & 250 \\
 & Mistral-7B-v0.3 & self-consistency & 0.526 & 250 \\
 & Mistral-7B-v0.3 & QLoRA\textsuperscript{a} & 0.566 & 250 \\
 & gpt-5.2\textsuperscript{c} & zero-shot & 0.721 & 250 \\
\midrule
stance toward India & two coders &  & 0.521 & 250 \\
 & Llama-3.1-8B & self-consistency & 0.407 & 250 \\
 & gpt-5.2\textsuperscript{c} & zero-shot & 0.664 & 250 \\
\midrule
stance toward Pakistan & two coders &  & 0.557 & 250 \\
 & Llama-3.1-8B & zero-shot & 0.455 & 250 \\
 & gpt-5.2\textsuperscript{c} & zero-shot & 0.775 & 250 \\
\midrule
target type & two coders &  & 0.682 & 500 \\
 & Perspective & rule on scores & 0.228 & 500 \\
 & DeBERTa-v3 & NLI zero-shot & 0.377 & 500 \\
 & Gemma-3-12B & zero-shot & 0.416 & 500 \\
 & Gemma-3-12B & QLoRA\textsuperscript{b} & 0.480 & 100 \\
 & RoBERTa-large & fine-tuned\textsuperscript{a} & 0.487 & 500 \\
\bottomrule
\end{tabularx}
\vspace{2pt}
{\footnotesize \textsuperscript{a}Five-fold cross-validation.
\textsuperscript{b}A single 80/20 split, scored on the 100 held-out items.
\textsuperscript{c}Voted on the reference's disputed items.\par}
\end{table}

When we score the production labels against the items on which both coders agreed, they reach $\kappa = 0.775$ on narrative frame, 0.799 on sentiment, and 0.825 and 0.830 on stance toward India and toward Pakistan (Table~\ref{tab:reliability}). Those items are easier than the disputed ones, so these figures are upper bounds. On narrative frame, where a reviewer resolved every disputed item, the labels reach 0.707 against the full reference; on sentiment and stance, the full reference includes gpt-5.2's own votes, so we do not score the labeler against it. The model agrees more with the coder who worked from translations ($\kappa$ 0.654 to 0.739 on frame, sentiment, and the two stance axes) than with the coder who reads Hindi and Urdu (0.541 to 0.595). This pattern suggests that the model's labels, like that coder's, lack regional context. We share our labels for every comment and post, keyed by Reddit ID, with reviewers through an anonymized repository\footnote{\url{https://anonymous.4open.science/r/IndoPakRedditLabels/}}, and we will make it public on publication.

\subsection{Ethical considerations}
\label{sec:data-ethics}

Our institution's review board determined that this study is not human-subjects research, because we analyze public posts and comments and do not interact with the people who wrote them. We quote comments verbatim, with their community, date, and score, because the wording and reception are our evidence, but we do not quote usernames, and we checked that no quoted comment names or identifies a private individual.

\section{Findings}
\label{sec:results}

This section discusses the results for our three research questions.
For RQ1 we work at the level of the comment, asking what these communities contested. For RQ2, we work at the community level within an event, asking why their contests differed; for RQ3, we follow participants to the reply dyad and then to the author, across community boundaries.

\subsection{RQ1: Contestation tracks frame and community, not toxicity}
\label{sec:results-contestation}

Comparing contestation across toxicity, frame, and community, we find that contestation tracks the community a comment is posted in and the narrative frame it uses (e.g., humanitarian or security framing), but hardly at all whether the comment is toxic.
Of the 297,025 comments with a frame and a toxicity label, 6.90\% carry the \texttt{controversiality} flag. The contestation rate rises to 17.43\% for humanitarian-framed comments in the three general-audience communities ($N = 4{,}005$, Wilson 95\% CI [16.28, 18.63]) and to 18.54\% for the same combination inside the crisis windows.

\begin{figure}[!h]
  \centering
  \includegraphics[width=.9\linewidth,alt={Contestation rate by community.}]{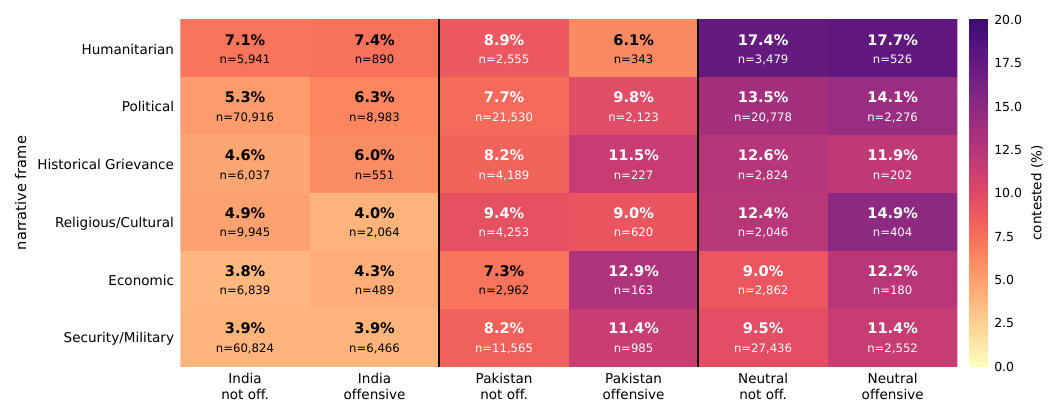}
  \caption{Observed contestation rate by community alignment, narrative frame, and offensiveness. General-audience communities are labeled \emph{neutral}.}
  \label{fig:r8heatmap}
\end{figure}

Figure~\ref{fig:r8heatmap} shows the contestation rate across community alignment, narrative frame, and offensiveness (targeted toxicity). Within each alignment group, the non-offensive (\textit{not off.}) and offensive columns track closely together, while rates differ substantially between groups. Toxicity accounts for the smallest gap: comments carrying an offensive label are contested 7.57\% of the time versus 6.82\% for non-offensive comments, a ratio of 1.11. By comparison, community alignment widens the contestation rate 2.47-fold, from 11.74\% in general-audience (\textit{Neutral}) communities through 8.24\% in \texttt{r/pakistan} to 4.75\% in India-aligned communities, and narrative frame widens it 1.77-fold, from 10.45\% under humanitarian framing to 5.89\% under economic framing. The feature that automated moderation acts on moves peer contestation the least. Holding frame constant, we find that, among humanitarian-framed comments, offensive ones are contested 10.23\% of the time ($n = 1{,}759$) and non-offensive ones 10.48\% ($n = 11{,}975$), a slight gap in the opposite direction from what an account based on tone would predict.

\begin{table}[!h]
\centering
\caption{Effect sizes for contestation by community, narrative frame, and toxicity (top), and the toxic-versus-non-toxic contrast within each frame (bottom).}
\label{tab:r8equiv}
\setlength{\tabcolsep}{5pt}
\footnotesize
\begin{tabularx}{\linewidth}{@{}>{\raggedright\arraybackslash}Xrrr@{}}
\toprule
\multicolumn{4}{@{}l}{\emph{Effect sizes (contestation $\sim$ factor)}} \\
factor & rate ratio & Cramer's $V$ & interpretation \\
\midrule
community    & 2.47$\times$ & 0.114 & weak \\
frame        & 1.77$\times$ & 0.040 & negligible \\
offensiveness & 1.12$\times$ & 0.009 & negligible \\
\bottomrule
\end{tabularx}

\vspace{4pt}

\begin{tabularx}{\linewidth}{@{}>{\raggedright\arraybackslash}Xrrrr@{}}
\toprule
\multicolumn{5}{@{}l}{\emph{Within-frame offensive vs.\ not ($\Delta$ = offensive $-$ not)}} \\
frame & $\Delta$ (pp) & 95\% CI (pp) & $\phi$ & TOST \\
\midrule
Humanitarian         & $-0.29$ & $[-1.80, +1.21]$ & 0.003 & equiv. \\
Historical Grievance & $+0.74$ & $[-1.02, +2.50]$ & 0.007 & n.s.\ \\
Political            & $+1.00$ & $[+0.51, +1.49]$ & 0.012 & equiv. \\
Religious/Cultural   & $-0.52$ & $[-1.47, +0.43]$ & 0.007 & equiv. \\
Security/Military    & $+0.60$ & $[+0.10, +1.11]$ & 0.007 & equiv. \\
Economic             & $+1.85$ & $[+0.01, +3.69]$ & 0.019 & n.s.\ \\
\bottomrule
\end{tabularx}

\vspace{2pt}
{\footnotesize ``equiv.'' marks a frame where the $\pm 2$pp equivalence test
rejects at $p < 0.05$. The two ``n.s.'' frames are under-powered for equivalence
(small offensive $n$ widens the CI), not evidence of a large gap: both have
$\phi < 0.02$.}
\end{table}

Since the corpus is large enough that a chi-square test flags trivial gaps as significant, we judge these contrasts by effect size and equivalence rather than by $p$ alone. Table~\ref{tab:r8equiv} reports the per-frame equivalence tests and the three effect sizes in full. The within-frame offensive-versus-not gap is statistically equivalent within a $\pm 2$pp margin (two one-sided tests~\cite{lakens2017equivalence}, $p < .05$) in four of the six frames, and its effect size is negligible in every frame ($\phi < 0.02$); even the most significant gap, \textit{Political}'s ($p = 2.5\times10^{-5}$), still carries only $\phi = 0.012$. Applying the same effect-size ruler to the factors we do treat as large, all three associations are small in absolute terms, but their rank is clear: community ($V = 0.114$) above narrative frame ($V = 0.040$) above offensiveness ($V = 0.009$). Contestation is more strongly organized by context than by tone, although no single factor determines it.

\begin{figure}[!h]
  \centering
  \includegraphics[width=0.75\linewidth,alt={Gain in McFadden's pseudo-$R^2$.}]{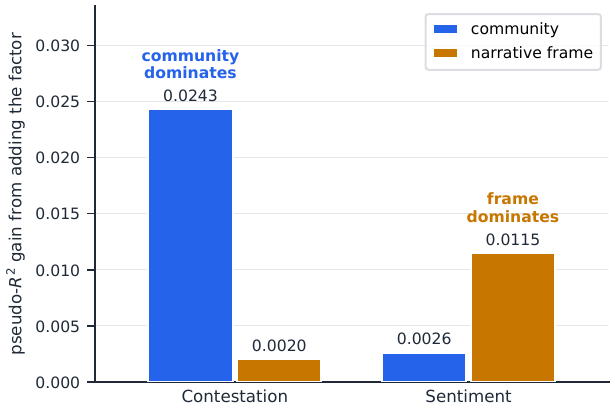}
  \caption{Gain in McFadden's pseudo-$R^2$ from adding community or narrative frame to a model of contestation (left) or sentiment (right) that already holds the other factor, crisis type, and content type, fitted on crisis-window comments in the ten communities. A taller bar means the factor adds more to the fit.}
  \label{fig:crossover}
\end{figure}

The small effect of frame on contestation did not mean that frame mattered little. We compared the two factors using the gain in McFadden's pseudo-$R^2$~\cite{mcfadden1974conditional} from adding each to a model that already included the other, on comments in the crisis windows (Figure~\ref{fig:crossover}). For contestation, adding community raised the pseudo-$R^2$ by 0.024 and adding frame by 0.002. For sentiment, the order reversed: frame added 0.012 and community 0.003. Community thus organized what a community contested, and frame organized the affect in which a comment was expressed.

\begin{figure}[!h]
  \centering
  \includegraphics[width=.8\linewidth,alt={Contestation rate per community.}]{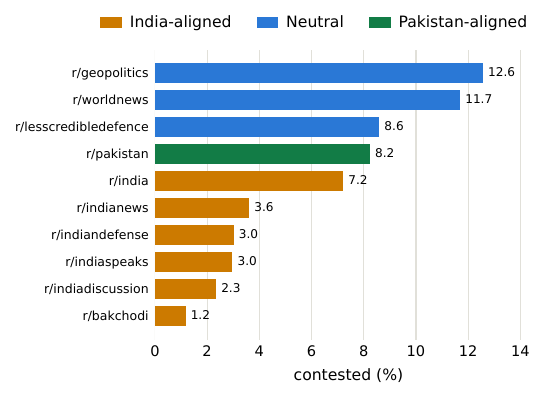}
  \caption{Contestation rate per community, coloured by alignment group.
  General-audience communities are labeled \emph{neutral}.}
  \Description{Horizontal bar chart of controversiality rate for the ten
  communities in the three alignment groups. The three general-audience
  communities occupy the top three positions at 12.6, 11.7 and 8.6 percent,
  followed by the Pakistan-aligned community at 8.2 percent, then the six
  India-aligned communities from 7.2 percent down to 1.2 percent. No
  India-aligned community except r/india exceeds any general-audience one.}
  \label{fig:r8subreddits}
\end{figure}

Figure~\ref{fig:r8subreddits} shows that the community effect does not come from a single community. All three general-audience communities ranked above every other community: \texttt{r/geopolitics} at 12.57\%, \texttt{r/worldnews} at 11.69\%, and \texttt{r/lesscredibledefence} at 8.59\%. The most contested India-aligned community, \texttt{r/india}, reached 7.21\%. When we refit the model with a fixed effect for each subreddit, which estimates the frame gradient net of the community a comment sits in, the model fit better than the alignment model (AIC 142,208.9 against 144,180.0), and the frame gradient held: the humanitarian-to-economic odds ratio fell only from 1.67 to 1.57. Net of subreddit, a toxic comment had 1.17 times the odds of being flagged (95\% CI [1.12, 1.23], $p = 3.4 \times 10^{-11}$), a smaller association than the humanitarian-to-economic contrast. The ordering also held without Operation Sindoor ($N = 228{,}386$): general-audience communities were flagged at 13.79\%, \texttt{r/pakistan} at 8.23\%, and India-aligned communities at 4.96\%, and humanitarian framing at 11.22\% against 6.21\% for economic framing. We reviewed all 698 flagged humanitarian comments in general-audience communities; 86.7\% had no toxicity label, and 11.3\% included a link. We read 479 comments that were between 15 and 120 words. What recurred were disputed factual claims about civilian harm and rights: ratios of troops to residents, counts of detainees, communications blackouts, and comparisons with other contested territories. 

One flagged comment put the comparison bluntly: ``India is literally doing the same thing in Kashmir that china is doing in Xianjing.'' Another attacked a proposed remedy rather than an opponent and closed on a sarcasm marker: ``Surely creating or exacerbating humanitarian disasters will show the military regime in Pakistan whats up! We all know governmental elites are the first to feel the pain /s.'' Neither comment insults anyone, and both drew votes in both directions. These communities divided over whether a claim was true, not over whether it was civil. The same dispute over facts appeared in other combinations. A flagged humanitarian comment in an India-aligned community turned the claim on the community's own army: ``There has been proof of the worst crimes by the army against Kashmiris like killing kids, sexually assaulting women,etc. So this isn't a big surprise'' In general-audience communities, flagged comments under historical-grievance framing disputed which side had acted first, as in ``India annexed Kashmir and killed civilians first, if you want to keep back tracing the causes\ldots''

Three caveats qualify these results. First, \texttt{controversiality} tracks low net score: 23.7\% of comments at a net score of zero or below carry the flag, against 0.02\% of those above one hundred. The flag therefore registers contested reception rather than observed deliberation, and we avoid deliberation language throughout. Second, general-audience communities draw a more transient audience, at 2.69 comments per author against 4.51 in India-aligned communities and 5.04 in \texttt{r/pakistan}. A transient audience could split votes on its own, so we test that alternative rather than argue against it. Adding a control for author activity moves the general-audience odds ratio only from 2.785 to 2.697. Restricting to committed participants widens rather than narrows the gap, from a rate ratio of 2.47 across all authors to 2.81 among those with at least twenty comments. A transient-voter artifact would have moved both the other way. Third, alignment and community are collinear: the fixed-effects result shows that these three general-audience communities are contested more, a descriptive claim rather than one about general-audience communities as a type.

\begin{figure}[!h]
  \centering
  \includegraphics[width=0.98\linewidth,alt={Comparision witing each narrative frame to show how contested the frame is.}]{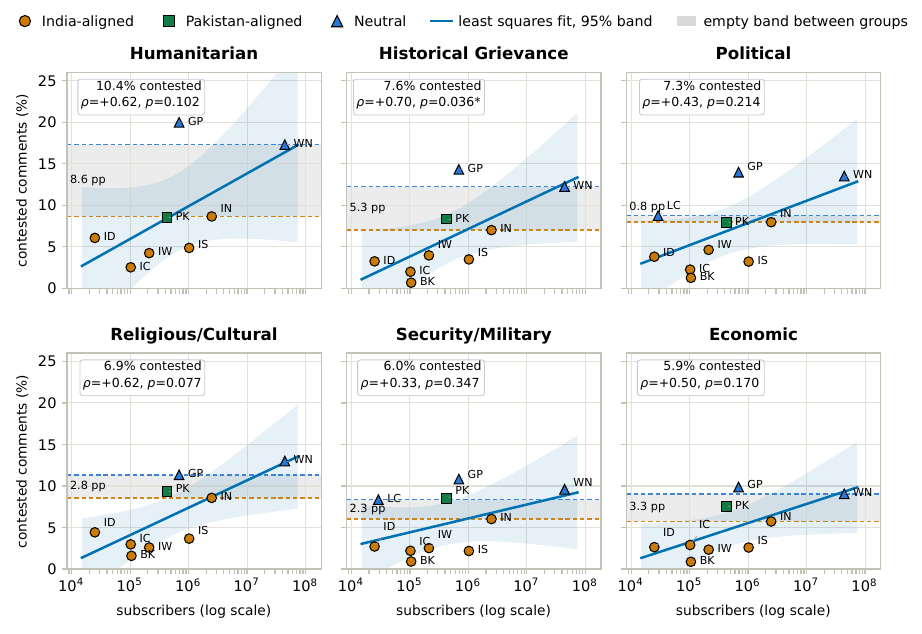}
  \caption{Comparison within each narrative frame, with panels ordered
  by how contested the frame is overall. The shaded horizontal band is the empty
  range between the highest India-aligned community and the lowest
  general-audience one, annotated with its width in percentage points. One point
  per community per frame; combinations resting on fewer than one hundred comments are
  omitted, which is why some panels carry fewer than ten points. An asterisk
  marks $p < .05$ uncorrected, across six panels.
  Community codes: IN~\texttt{r/india}, IS~\texttt{r/IndiaSpeaks},
  ID~\texttt{r/IndianDefense}, IC~\texttt{r/indiadiscussion},
  IW~\texttt{r/indianews}, BK~\texttt{r/bakchodi}, PK~\texttt{r/pakistan},
  WN~\texttt{r/worldnews}, GP~\texttt{r/geopolitics},
  LC~\texttt{r/lesscredibledefence}.}
  \Description{Six scatterplots, one per narrative frame, ordered by how
  contested the frame is overall. Each plots contestation rate against
  subscriber count on a logarithmic horizontal axis, with one labelled point per
  community, a least squares fit and a 95 percent band. A shaded horizontal band
  marks the empty range between the highest India-aligned community and the
  lowest general-audience one, annotated with its width in percentage points.
  That band is non-empty in all six panels, from 0.8 points under political
  framing to 8.6 under humanitarian framing. Only historical-grievance framing
  reaches a significant rank correlation with subscriber count, one panel of six,
  uncorrected.}
  \label{fig:r8sizeframe}
\end{figure}

Figure~\ref{fig:r8sizeframe} tests audience size against alignment by plotting each community's contestation rate against its subscriber count within each frame. Only historical-grievance framing correlates with subscriber count at a conventional threshold ($\rho = 0.700$, $p = .036$), one panel of six, and the correlation does not survive correction across the six. In the other five frames, the rank correlation is positive but not significant ($\rho$ from 0.33 to 0.62), so a community's size does not predict how much it contests a given frame. What every panel shares is the alignment gap: the shaded band separates every general-audience community from every India-aligned one, by 0.8 percentage points under political framing up to 8.6 under humanitarian framing. Ten communities is a thin basis for a correlation, so we report this as a robustness check of the alignment result rather than as a finding of its own.

\subsection{RQ2: The frame mix tracks crisis type, and its sentiment varies by community}
\label{sec:results-rq2}

\subsubsection{An event's signature tracks crisis type, not year}
Comparing the six events' signatures, we find that the mix of frames a crisis draws, and the sentiment attached to each, tracks crisis type more closely than year. We represent each event, pooled across communities, as a matrix of joint proportions of six frames by three sentiments (its signature), and compare events by cosine distance, with 95\% intervals from a bootstrap over threads. Pulwama and Operation Sindoor took place six years apart, and both were military events. Their signatures are almost identical (cosine distance 0.0074 [0.0022, 0.0313]). Pulwama and the constitutional change six months later sit far apart (0.2058 [0.1359, 0.2668]). The separation holds across all pairs: the largest upper bound among the six military pairs is 0.0768, and the smallest lower bound among the four military-to-constitutional pairs is 0.1359, a gap of 0.0591. A permutation test of event labels finds that every pair of signatures differs, including the military pairs, showing that events are distinguishable, not that military events cluster. The 2021 ceasefire does not fit the pattern. It sits closer to the constitutional change (0.0377) than several military events sit to one another, so the regularity holds only for the contrast between military and constitutional events.

Table~\ref{tab:signature-metrics} in Appendix~\ref{app:signature} repeats the comparison under two other distances, because a signature is a probability distribution and cosine distance treats it as a vector. We measure every pair of events with the Jensen-Shannon divergence~\cite {lin1991divergence} and Euclidean distance. Interestingly, changing the metric does not change the separation: under every metric, each military pair sits closer than each pair of a military event and the constitutional change, and the two families' intervals do not overlap (gaps of 0.0127 under the Jensen-Shannon divergence and 0.0533 under Euclidean distance). Frame composition divides sharply by crisis type ($\chi^2(10) = 17{,}743.3$, Cram\'er's $V = 0.235$, $N = 160{,}050$). Adjusted residuals locate the division, with security and military framing concentrated in military events ($z = +129.19$) and political framing in the constitutional change ($z = +101.22$). Community alignment moderates this composition. A nested likelihood-ratio test of the interaction between crisis type and community returns $\chi^2(20) = 834.4$ ($p = 6.92 \times 10^{-164}$), with standard errors clustered on 37,762 authors, although it raises McFadden's pseudo-$R^2$ only from 0.0523 to 0.0544. Thus, the frames a community reaches for track crisis type above all, while each alignment group shifts that mix by a small but reliable amount.

\begin{figure}[h!]
  \centering
  \includegraphics[width=.85\linewidth,alt={Divergence between the sentiment of communities by narrative frame and crisis event.}]{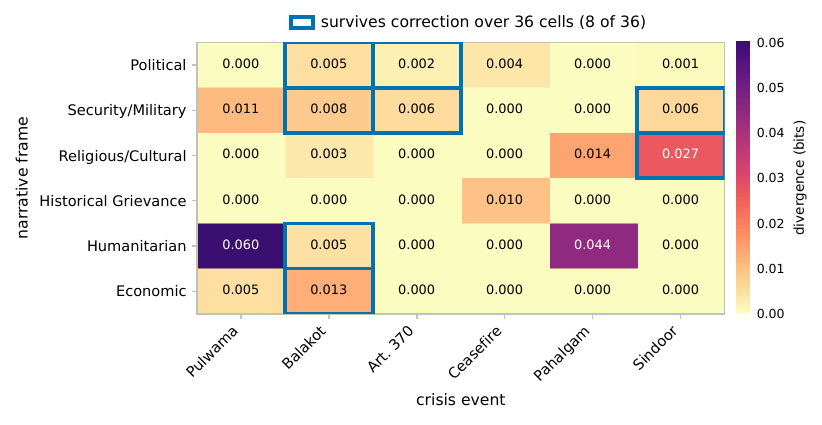}
  \caption{HMiller--Madow-corrected Jensen--Shannon divergence (bits) between the sentiment distributions of India-aligned communities and \texttt{r/pakistan}, by narrative frame and crisis event. Outlined combinations survive Benjamini--Hochberg correction across all 36.}
  \Description{Heatmap of cross-community sentiment divergence for six narrative frames against six crisis events. Shade encodes magnitude and an outline marks the eight combinations of thirty-six that survive correction. The largest value, 0.060 for humanitarian framing during Pulwama, is not outlined, because it rests on few comments.}
  \label{fig:r1jsd}
\end{figure}

Figure~\ref{fig:r1jsd} maps where, within an event, the two aligned sides diverge in sentiment, and they diverge inside particular frames, not across the board. To locate these divergences, we use the Jensen-Shannon divergence for a second comparison, between communities instead of between events. For each of the 36 frame-event combinations, we compare the sentiment distribution of the India-aligned communities with that of \texttt{r/pakistan}. We correct each divergence for small-sample bias with the Miller-Madow estimator~\cite{miller1955note,paninski2003entropy} and test it against an exact hypergeometric null, with Benjamini-Hochberg correction across the 36 combinations~\cite{benjamini1995controlling}. Eight of the 36 combinations survive. The largest is religious and cultural framing during Operation Sindoor (0.0266, adjusted $p = .006$), followed by economic framing (0.0128) and security and military framing (0.0082) during Balakot. Pooling events by their role sharpens the contrast: humanitarian framing diverges 26 times more during the attacks that opened a crisis (0.0413, adjusted $p = .018$) than during the retaliations that followed them (0.0016, adjusted $p = .036$).

The largest divergent combination shows what the divergence consists of. Religious and cultural framing during Operation Sindoor covers 2,013 comments across the India-aligned communities and \texttt{r/pakistan}. Our labeler, gpt-5.2, reads 79.3\% of the India-aligned comments in this combination as negative against 63.6\% of those in \texttt{r/pakistan}, and 8.3\% as positive against 21.6\%. India-aligned comments mostly characterize the other side's religiosity and attribute its positions to religious identity or to religious schooling, as in the highest- and fourth-highest-scoring of the 1,410 negative India-aligned comments in this combination (\texttt{r/india}, May 10, 2025, score 2,133; \texttt{r/IndiaSpeaks}, April 29, 2025, score 815):
\begin{quote}
``When your point of view is only informed by your religious identity, then there is very literal chance of talking any sense.''\\[3pt]
``This is what state funded Madarsa are teaching to young children in the name of education.''
\end{quote}
Comments in \texttt{r/pakistan} carry a register with almost no counterpart on the other side. They quote scripture directly, as in ``Hasbunallahu wa Ni'mal Wakeel'', which the comment glosses as ``Sufficient for us is Allah, and He is the best disposer of affairs'' (May 10, 2025, score 76), and credit the strikes to divine help (May 11, 2025, score 40):
\begin{quote}
``Every missile that we fired left Pakistan with ALLAH HU AKBAR. Truly this help was from ALLAH and ALLAH alone.''
\end{quote}
A few positive comments argue instead against generalizing about Hindus: ``Seeing Pakistani Hindus supporting Pakistan on X in this hard time, we shouldn't use generalized terms'' (May 10, 2025, score 60). All three rank among the six highest-scoring of the 51 positive \texttt{r/pakistan} comments in this combination. This result suggests that the positive tail the divergence statistic detects is largely devotional, and that one frame label can cover two different speech acts rather than one disputed topic.

\begin{figure}[h!]
  \centering
  \includegraphics[width=0.8\columnwidth,alt={Share of comments labeled negative in aeach crisis window.}]{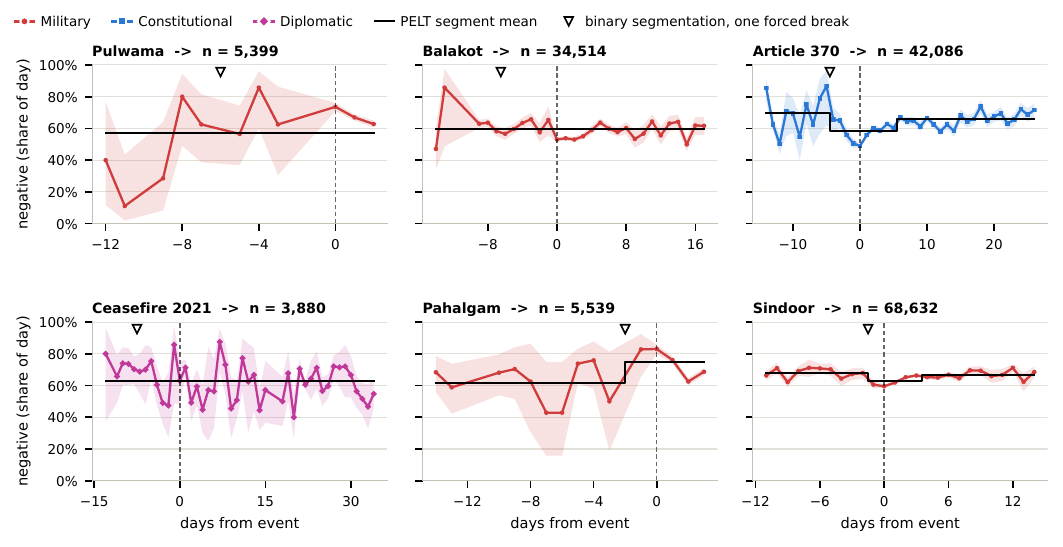}
  \caption{Daily share of comments labeled negative in each crisis window, with
  95\% Wilson intervals, colored by crisis type. Days with fewer than five
  comments are omitted, and the dashed line marks the event date. Black steps show the mean of each segment that PELT finds with a BIC penalty. 
  An open triangle marks the single change point that binary segmentation places when we set it to find exactly one per event.}
  \label{fig:temporal}
\end{figure}

Figure~\ref{fig:temporal} shows the daily share of comments labeled negative for each crisis window. Crises with more comments have less volatile daily sentiment (Spearman $\rho = -0.77$, $p = .072$, $n = 6$), and the correlation is the same with three-, five-, and seven-day windows. We also search each event's daily series for change points with PELT~\cite{killick2012pelt}, adding a change point only when it lowers the fitting cost by more than a penalty, and we set the penalty using the Bayesian information criterion~\cite{schwarz1978estimating}. PELT places change points in three events (the revocation of Article 370, Operation Sindoor, and the Pahalgam attack), all within six days of the event date. Around the changes to Article 370, the negative share falls from 69.6\% to 58.4\% between four days before the event date and five days after it, then returns to 65.7\%. Operation Sindoor shows a smaller dip, from 68.0\% to 62.7\% between the day before the event date and three days after it, before returning to 66.8\%. The Pahalgam attack goes the other way, with the negative share rising from 61.3\% to 74.6\% from the day before the event date. PELT places none in Pulwama, Balakot, or the 2021 ceasefire. Binary segmentation~\cite{truong2020selective}, which returns as many breaks as events, places one in every event, including the three where PELT finds none. Interestingly, the constitutional change, the most politically contested event in the corpus, is the event on which communities converge most in sentiment.

\subsubsection{Aligned communities read the same frame more positively}
\label{sec:results-affect}

Frame organized sentiment far more than community did, and the same frame still carried different affect in different communities. We modeled sentiment as a multinomial logit~\cite{king2000interpretation} on narrative frame and community, controlling for crisis type and content type, with standard errors clustered on authors. Adding frame to a model that held community raised McFadden's pseudo-$R^2$ from 0.0397 to 0.0513 ($\chi^2(10) = 3{,}189.8$), the mirror image of contestation, where community dominated. Adding the interaction between frame and community raised it from 0.0513 to 0.0520 ($\chi^2(20) = 179.8$, $p = 1.05 \times 10^{-27}$, $N = 160{,}050$). The interaction is reliable but small, about a sixteenth of what the frame adds, and it held when we entered content type only as opinion against everything else ($\chi^2(20) = 186.0$). We expected the difference between communities to be negative, but it was not. It appeared positive: aligned communities rendered the frames that engaged their side as heroic rather than grim.

\begin{figure}[h!]
  \centering
  \includegraphics[width=\linewidth, alt{Share of comments read as negative vs. positive for each narrative frame.}]{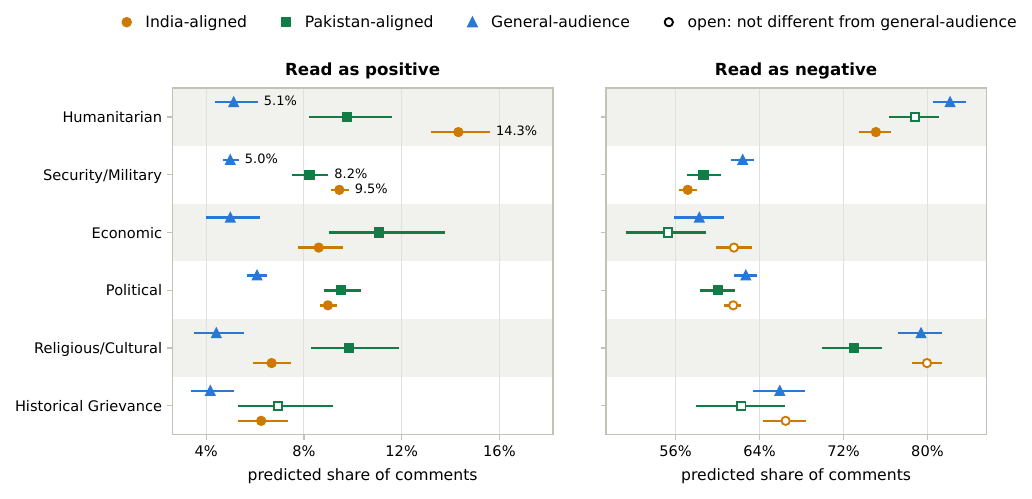}
  \caption{Predicted share of comments read as positive (left) and as negative (right), by narrative frame and alignment group, from the multinomial logit, with 95\% simulation intervals. General-audience communities are the reference: an open marker is an aligned group that does not differ from them after Holm correction.}
  \Description{Two panels of point estimates with 95 percent simulation
  intervals, one for probability of negative sentiment and one for probability of
  positive sentiment, across six narrative frames sorted by the size of the
  general-audience gap. Each frame carries two estimates, the Pakistan-aligned
  minus India-aligned gap and the general-audience minus India-aligned gap. A
  filled marker is an interval that excludes zero and an open marker one that
  spans it. In the positive panel every general-audience estimate falls below
  zero with a filled marker, the largest being humanitarian framing at about
  minus nine percentage points. In the negative panel the general-audience
  estimates for security and military framing and for humanitarian framing sit
  above zero.}
  \label{fig:r5ame}
\end{figure}

Figure~\ref{fig:r5ame} shows the difference, with predicted shares and simulation intervals computed from the multinomial logit. Under security and military framing, the multinomial logit predicts that 9.4\% of India-aligned comments read as positive, against 8.2\% in \texttt{r/pakistan} and 5.0\% in general-audience communities. Under humanitarian framing the India-aligned share rises to 14.3\%, against 5.1\% in general-audience communities. India-aligned communities read every frame more positively than general-audience communities by 2.1 to 9.2 percentage points, and every gap survives Holm correction~\cite{holm1979simple} across the 24 contrasts in the figure. \texttt{r/pakistan} does so in five of the six frames. Its gap under historical grievance (2.8 points) does not survive the correction. The India-aligned gap is widest under humanitarian framing. The negative side of the same model shows where the contrast comes from: under security and military framing, general-audience comments are 5.2 percentage points more likely than India-aligned ones to read as negative, while \texttt{r/pakistan} does not differ from the India-aligned communities. This result suggests that general-audience detachment, not opposition between the two aligned sides, carries the gap, and that what sets the aligned communities apart is their engagement with the frame more than their position on the conflict.

Since sentiment measures stance indirectly, we tested it by refitting the model with both stance axes as controls. The interaction strengthened rather than weakened ($\chi^2(20) = 287.6$, $p = 2.75 \times 10^{-49}$; 299.3 with content type entered as opinion against the rest), and several contrasts that had been indistinguishable became significant, so affect and stance are separable in these data. Two combinations did not survive the control. The \texttt{r/pakistan} gap in negativity under humanitarian framing reversed and lost significance, moving from $+0.038$ to $-0.018$, and the corresponding gap under economic framing shrank by 38\%. Security and military framing show the mechanism. We count in-group pride markers among the 5,208 positively read security and military comments inside the crisis windows. At least one marker appeared in 52.2\% of India-aligned comments ($n = 3{,}698$) and 47.6\% of those in \texttt{r/pakistan} ($n = 517$), against 33.8\% of general-audience comments ($n = 993$). First-person plural forms alone appeared in 40.5\% of India-aligned comments and 36.8\% of those in \texttt{r/pakistan}, against 22.4\% of general-audience comments. India-aligned comments under this frame claim the armed forces in the first person plural, as in ``the fact that our forces were able to target it effectively in a single strike is significant''. General-audience comments attached their positive sentiment to de-escalation instead: ``Two nuclear powers deciding to not escalate is certainly a victory for everyone involved.'' The two communities were not disagreeing about the military. They were performing different speech acts under one frame label, which is the qualitative form of the gap in Figure~\ref{fig:r5ame}. We offer the marker counts as corroboration rather than as an estimate, because the lexicon is a coarse heuristic built for precision.

The interaction held under every check we ran. It strengthened without Operation Sindoor, the largest single block of comments ($\chi^2(20) = 231.0$,
$N = 91{,}418$), and held when we kept only labels in which the labeler reported high confidence ($\chi^2(20) = 176.1$, $N = 88{,}534$) and when we extended the model to the full corpus with crisis phase controlled ($\chi^2(20) = 533.6$, $N = 305{,}165$). It also held within content types: among opinion comments the test gave $\chi^2(20) = 161.6$ ($p = 3.8 \times 10^{-24}$, $n = 125{,}107$), and among all other comments $\chi^2(20) = 46.5$ ($p < .001$, $n = 34{,}941$), with a similar gain in fit in both. The difference therefore reflects how these communities expressed a position, not the mix of genres they posted.

\subsubsection{Communities reward negative security framing differently}
\label{sec:results-reward}

The communities also differed in how they reinforced affect, although only part of that difference held up. We regressed a sign-preserving log of comment score on sentiment interacted with frame and community, controlling for crisis type, crisis phase, content type, and whether a comment opened a thread, with standard errors clustered on authors.  Negative comments were reinforced least in \texttt{r/pakistan}, at 1.239 against 1.418 in India-aligned communities (a difference of $-0.179$, 95\% interval $[-0.246, -0.116]$), and this held without Operation Sindoor (1.173 against 1.470, a difference of $-0.296$, $[-0.380, -0.213]$). Neutral comments followed the same order in both samples. The general-audience result did not hold. With every event included, general-audience communities reinforced negative comments most (1.566, or $+0.148$ against India-aligned communities, $[0.081, 0.210]$), but without Operation Sindoor they validated them less than India-aligned communities did (1.335, or $-0.134$, $[-0.217, -0.054]$). Positive comments showed no reliable difference with every event included, while without Operation Sindoor both \texttt{r/pakistan} and general-audience communities validated them less than India-aligned communities did ($-0.272$ and $-0.410$). We therefore treat the lower validation for negative comments in \texttt{r/pakistan} as the robust result, and the general-audience premium as dependent on a single event.

\begin{figure}[h!]
  \centering
  \includegraphics[width=\linewidth,alt={Percentage of security and military comments in crisis windows.}]{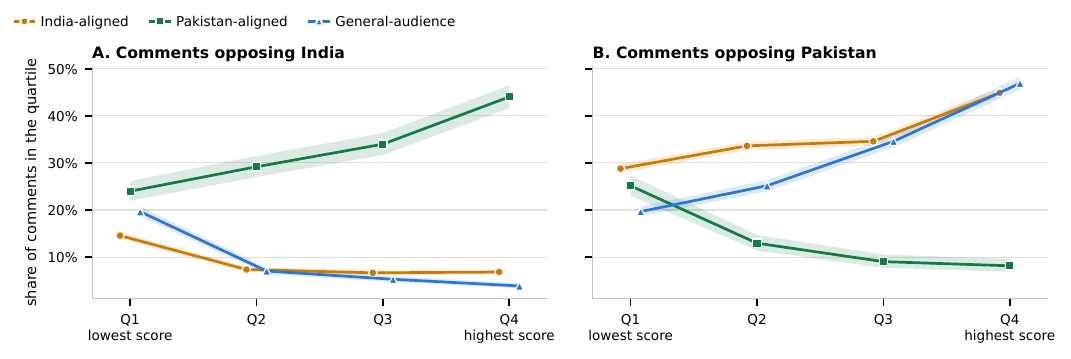}
  \caption{Share of security and military comments in the crisis windows that
  oppose India (left) and Pakistan (right), by score quartile within each
  community, with 95\% Wilson intervals.}
  \Description{Two line charts with four score quartiles on the horizontal axis.
  Left, comments opposing India: the share falls from about 15 to 7 percent in
  India-aligned communities and from about 20 to 4 percent in general-audience
  communities, and rises from 24 to 44 percent in the Pakistan-aligned community.
  Right, comments opposing Pakistan: the share rises from about 29 to 45 percent
  in India-aligned communities and from 20 to 47 percent in general-audience
  communities, and falls from 25 to 8 percent in the Pakistan-aligned community.}
  \label{fig:r5stance}
\end{figure}

Three properties of the score limit what this result can show. We used a sign-preserving log because 9.25\% of scores in the corpus are negative, which rules out a count model. Score is a noisy proxy for endorsement, and it draws on the same votes as our contestation measure, so the reward and contestation results are two readings of one platform signal, not independent tests. Most importantly, the model does not control for stance, and stance differed sharply across the score range (Figure~\ref{fig:r5stance}). Among security and military comments, the share opposing India fell from the lowest to the highest score quartile in India-aligned communities (from 14.6\% to 6.8\%) and in general-audience communities (from 19.6\% to 3.9\%), and rose in \texttt{r/pakistan} (from 24.0\% to 44.0\%). The share opposing Pakistan moved the other way: it rose in India-aligned communities (from 28.8\% to 44.9\%) and in general-audience communities (from 19.7\% to 46.9\%), and fell in \texttt{r/pakistan} (from 25.1\% to 8.1\%). The highest-scoring security comments in general-audience communities were thus far more likely to oppose Pakistan than India. One of the lowest-scoring states a position without insulting anyone: ``Maybe India just doesn't get it after losing 5 aircraft in a failed operation.'' Part of what separates these communities may therefore be which stances they rewarded rather than how they received a fixed comment, and our controls do not remove that.

\subsection{RQ3: Crossing the boundary sharpens opposition}
\label{sec:results-rq3}

\subsubsection{Threads mix frames, replies keep them, and frame-crossing replies are more toxic}
\label{sec:results-threads}

Following conversation from threads to reply pairs and then across community boundaries, we find that crossing a boundary sharpens opposition rather than softening it. We expected conversation to split into frame-organized pockets, but it does not. Figure~\ref{fig:app-cooccurrence} in Appendix~\ref{app:cooccurrence} draws the frame co-occurrence graph of each event. Within threads, every pair of frames co-occurs in every event, so all six graphs are complete, with all fifteen frame pairs present. Louvain community detection~\cite{blondel2008fast} finds a modularity of 0.000 in every event and at every seed, across 4,210 threads, while graphs with the same edge weights shuffled at random have mean modularities between 0.005 and 0.045. The observed graphs are therefore no more separable than chance ($p = 1.000$ in every event). No frame-segregated pockets appear at the level of the thread.

Table~\ref{tab:persistence} reports a second pattern that coexists with this mixing: strong persistence from one reply to the next. Frame assortativity~\cite{newman2003mixing} across reply pairs ranges from 0.28 to 0.39 by event. We fit one-versus-rest logistic models of whether a reply keeps its parent's frame, over 95,301 reply pairs with standard errors clustered on 24,059 authors. Every frame persists above its base rate, and the least common frames persist most: a reply to an economic comment has 26.0 times the odds of staying economic, and the odds ratios are 20.2 for historical grievance and 18.3 for religious or cultural framing, against 7.3 for humanitarian, 6.4 for security and military, and 5.3 for political framing. All six survive Benjamini-Hochberg correction. Clustering on authors keeps the same participants' repeated replies from inflating the precision of these estimates, although it does not remove those replies from the estimates themselves. We call the pattern persistence rather than contagion, because the design accounts for topical coherence within a thread but not for who chooses to reply.

\begin{table}[!h]
\centering
\caption{Agreement between our two coders, and between each layer's deployed
labeller and the coders' labels.}
\label{tab:reliability}
\small
\setlength{\tabcolsep}{3pt}
\begin{tabularx}{\linewidth}{@{}>{\raggedright\arraybackslash}Xrrrcl r@{}}
\toprule
 & \multicolumn{4}{c}{two coders} & \multicolumn{2}{c}{deployed labeller} \\
\cmidrule(lr){2-5}\cmidrule(l){6-7}
layer & $n$ & raw & $\kappa$ & 95\% CI & system & $\kappa$ \\
\midrule
relevance & 200 & 0.913 & 0.818 & [0.711, 0.914] & gpt-5.2 & 0.893 \\
content type & 250 & 0.872 & \textbf{0.619} & [0.471, 0.748] & gpt-5.2 & 0.432 \\
narrative frame & 250 & 0.797 & \textbf{0.700} & [0.615, 0.778] & gpt-5.2 & 0.775 \\
sentiment & 250 & 0.733 & 0.541 & [0.426, 0.643] & gpt-5.2 & 0.799 \\
stance toward India & 250 & 0.717 & 0.535 & [0.429, 0.632] & gpt-5.2 & 0.825  \\
stance toward Pakistan & 250 & 0.783 & 0.579 & [0.466, 0.685] & gpt-5.2 & 0.830 \\
target type & 500 & 0.893 & \textbf{0.671} & [0.589, 0.753] & RoBERTa & 0.487\textsuperscript{b} \\
toxicity at 0.5 & 200 & 0.745 & 0.477 & [0.360, 0.588] & Perspective & 0.670\textsuperscript{a} \\
\bottomrule
\end{tabularx}
\vspace{2pt}
{\footnotesize \textsuperscript{a}The 0.5 threshold against the reviewer-resolved labels of all
200 items. \textsuperscript{b}Five-fold cross-validation on all 500 sampled
items; the deployed classifier was then trained on all of them.\par}
\end{table}

Replies that cross to a different frame are more toxic and shift sentiment less often. Pooled across events, cross-frame replies score 0.0284 higher on Perspective's toxicity score than same-frame replies (95\% CI $[0.0251, 0.0317]$, rank-biserial $r = 0.102$, Mann-Whitney $p = 3.44 \times 10^{-155}$, $N = 93{,}368$). The toxicity holds when we measure it as the share of replies at or above Perspective's 0.5 threshold rather than as a mean score ($+0.0130$), and the effect is small, so we report the rank-biserial $r$. Cross-frame replies also shift sentiment less often than same-frame replies. A security reply to a humanitarian parent in \texttt{r/india} replaces the claim about harm with a charge about terror camps (May 9, 2025, score 180): ``It is actually shameful that you guys are running terror camps in holy and
pious places like mosques.'' The same move appears in both directions. In \texttt{r/worldnews}, a parent describing the curfew in Kashmir (August 7, 2019, score 5) draws a religious counter-charge two days later (score 1):
\begin{quote}
\emph{Parent:} ``Theres over 750k military in Kashmir, with curfew, and death for
breaking the curfew. There is no telecom services for us to know how many
Kashmiris are being killed.''\\[3pt]
\emph{Reply:} ``What happened to the Hindus in your pakistan??? What do yiu have
to say about the exodus of kashmiri Pandits?? Fucking hypocrites and India haters
everywhere''
\end{quote}
In \texttt{r/india}, a parent defending the strikes on security grounds draws a
humanitarian reply the same day (May 6, 2025, scores 1 and 0):
\begin{quote}
\emph{Parent:} ``You want missiles to filter out kids in pok and somehow hit only
the terrorists?''\\[3pt]
\emph{Reply:} ``Let's talk again when that innocent kid is your brother/sister or
child. Eye for an eye makes the whole world blind, you ignorant fuck supporting a
jingoistic performative fascist and extremist government.''
\end{quote}
In each pair, the reply answers the frame with another frame instead of answering the claim. We speculate that this is why frame-crossing here reads as friction
rather than as argument that lands.

Two further patterns qualify these results. Frames diversify as threads deepen, with frame entropy rising by between 0.046 and 0.228 bits from top-level to deep comments, although the clustered intervals exclude zero in only four of the six events. The community difference in how strongly humanitarian framing persists is better read as a difference between venues than between alignment groups. Across the three groups, the spread in humanitarian self-transition is 0.088, while within the India-aligned group alone it reaches 0.160, from 0.056 in one community to 0.216 in another. That within-group spread rests on a community with only 36 humanitarian parent comments, so we report it as suggestive. The check needs at least two communities in a group, and \texttt{r/pakistan} is the only Pakistan-aligned community, so we cannot run it on that side.

We test both headline estimands across 36 specifications~\cite{simonsohn2020specification}, varying whether Operation Sindoor is included, the minimum confidence gpt-5.2 reported for its labels (none, 0.6, or 0.7), whether we keep only opinion comments, and, for the premium, whether toxicity is Perspective's score or its 0.5 cut. The cross-frame toxicity premium holds a median of 0.0241 across its 24 specifications (range $[0.0130, 0.0429]$). The historical-grievance persistence multiplier, the probability that a reply keeps its parent's frame divided by that frame's base rate, holds a median of 9.25 across its 12 specifications (range $[8.17, 10.15]$). Every interval excludes the null in both families.

\subsubsection{Crossing is incursion, and receiving communities penalize crossers}
\label{sec:results-crossing}

Profiling the authors who post outside their home group, we find no bridge-user type. We give each author a home group, the alignment group (India-aligned, Pakistan-aligned, general-audience, or diaspora) in which they comment most, and call a comment posted in any other group a crossing. If crossers acted as bridges, carrying a balanced or conciliatory register across the boundary, we would expect them to form a recognizable type, but they do not. We describe each of the 5,924 authors with at least ten comments by a thirteen-dimensional fingerprint, built from log-ratio coordinates~\cite{aitchison1982compositional,egozcue2003ilr} of the shares of their comments in each sentiment, stance, frame, and toxicity category. Density-based clustering (HDBSCAN~\cite{campello2013hdbscan}) of these fingerprints labels every author as noise at every setting we tried, and $k$-means silhouettes~\cite{rousseeuw1987silhouettes} stay near 0.1. The fingerprints form a continuum, not a set of kinds, and forcing clusters would manufacture types rather than find them. We therefore describe regions of the continuum. A Gaussian mixture with three components is stable under bootstrap resampling (adjusted Rand index~\cite{hubert1985comparing} 0.875), and we chose three on that stability, because information criteria kept falling through eight. The three regions hold 2,084, 2,061, and 1,779 authors, which we label Pakistan-sympathetic, India-partisan hawkish, and India-leaning civil.

Cross-community authors concentrate in the hawkish region. Region and the
corpus's cross-community flag are associated ($\chi^2(2) = 301.9$, Cram\'er's
$V = 0.226$, $p = 2.72 \times 10^{-66}$), with adjusted residuals of $+17.1$ for the India-partisan hawkish region and $-11.6$ for the Pakistan-sympathetic one. A continuous measure points the same way. We build two indices from the shares of each author's comments, each the mean of three components standardized across the 5,924 authors. The provocateur index of author $a$ is:
\begin{equation*}
  P_a = \tfrac{1}{3}\Bigl[\, z(t_a)
        + z\bigl(\max(h^{\mathrm{IN}}_a,\, h^{\mathrm{PK}}_a)\bigr)
        + z(n_a) \,\Bigr],
  \qquad z(x_a) = \frac{x_a - \bar{x}}{s_x},
\end{equation*}
where $t_a$ is the share of the author's comments with a targeted toxicity label, $h^{\mathrm{IN}}_a$ and $h^{\mathrm{PK}}_a$ are the shares labeled as opposing India and Pakistan, $n_a$ is the share labeled negative, and $\bar{x}$ and $s_x$ are the mean and standard deviation of a share across authors. Taking the larger of the two opposing share scores assigns an author to the side they oppose more, so the index does not depend on which country that is. We build the bridging index the same way from three other shares: the mean of the neutral-stance shares toward India and toward Pakistan, the share of comments that are not negative, and the share without a toxicity label. Neither index is a standard instrument: we built them to test the two readings of crossers that earlier work suggests, the bipartisan bridge~\cite{garimella2018echo} and the troll~\cite{cheng2017troll}. Authors without the flag score $+0.051$ on the bridging index and flagged authors $-0.026$, so cross-community participation and bridging are different behaviors here, and cross-community authors come disproportionately from the more combative end of the distribution.

\begin{figure}[h!]
  \centering
  \includegraphics[width=0.7\linewidth,alt={Share of comments opposing the out-group authors.}]{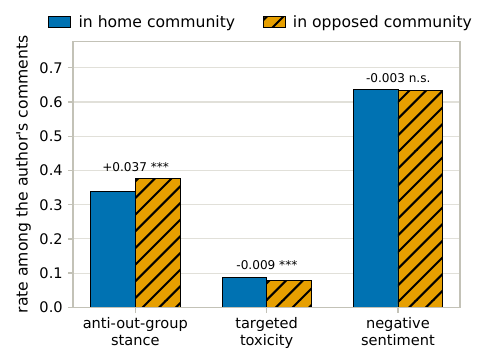}
  \caption{Share of comments opposing the out-group, carrying targeted toxicity, and read as negative, for the 1,560 authors with at least three comments both inside and outside their home group, each compared with themselves inside and outside it.}
  \Description{Grouped bar chart with three pairs of bars comparing in-group and
  out-group comments by the same 1,606 authors. Anti-out-group stance rises from
  0.339 to 0.376 when those authors post across the boundary, a significant
  increase. Targeted toxicity falls slightly from 0.088 to 0.079, also
  significant. Negative sentiment is unchanged at about 0.63.}
  \label{fig:r6raiding}
\end{figure}

Receiving communities penalize crossers, and the penalty does not single out the toxic ones. The 5,924 profiled authors post 16,266 comments outside their home group, and 49.4\% fall below the median score for comments in the same subreddit and crisis phase, which we count as penalized. Only 3,882 of these comments (23.9\%) are on the opposed-aligned side, from India-aligned authors in \texttt{r/pakistan} and from Pakistan-aligned authors in India-aligned communities. Most of the rest are aligned authors posting in general-audience communities (8,394) or general-audience authors posting in aligned ones (3,526). Relative to India-leaning civil authors, Pakistan-sympathetic authors are penalized at an odds ratio of 2.014 ($[1.673, 2.425]$, $p < .001$), while India-partisan hawkish authors are penalized no more than civil ones (1.033, $[0.871, 1.224]$, $p = .71$). The provocateur index predicts penalization (1.549, $[1.058, 2.266]$, $p = .024$), and so does the rate at which an author asks questions (2.146, $[1.062, 4.334]$, $p = .033$). The bridging index falls just short of significance (1.464, $[0.997, 2.148]$, $p = .052$). This result suggests that receiving communities penalize crossers who do not share their position more than they penalize toxicity, which means that `provoking' and `asking' both draw the penalty, while the hawkish region draws none. Figure~\ref{fig:r6raiding} follows the same authors across the boundary, and their conduct changes when they cross. We pair the 1,560 authors who commented at least three times both in their home group and outside it, and compare each author with themselves. Their share of comments opposing the out-group (Pakistan for India-aligned authors, India for Pakistan-aligned ones, and either country for general-audience authors) rises from 0.342 at home to 0.380 outside it (paired Wilcoxon $p < .001$). Their targeted toxicity falls slightly over the same move, from 0.088 to 0.079 ($p < .001$), so the same people push harder against the opposing position while their wording grows less toxic. However, negative sentiment does not change ($p = .852$).

Raising the minimum activity from five to ten to twenty comments moves the association between region and the cross-community flag from $V = 0.181$ to 0.226 and 0.228, and the provocateur odds ratio from 1.52 to 1.55 and 1.77 ($p = .024$, .024, and .013). Excluding Operation Sindoor leaves the provocateur effect at 1.632 ($[1.059, 2.516]$, $p = .027$). In other words, neither the activity threshold nor the largest crisis produces the provocateur penalty. Targeted toxicity is the least certain part of the fingerprint. Our RoBERTa-large classifier and a rule based on Perspective's attribute scores both decide only whether a toxic comment is targeted, and on that decision they agree at only $\kappa = 0.208$ across 30,382 toxic comments, so we do not treat their agreement as a robustness check.

\section{Discussion}

\subsection{Composition is the unit of analysis}

In our ten communities, contestation tracks where a comment is posted and what it claims far more than whether it carries a toxicity label. This gives a measured form to the argument that harm is a property of a comment in its community, which a score computed from the text discards by construction~\cite{berezin2025contextualharm,hartmann2026byebye}. It also adds the community to compositions of harm, which have so far combined attributes of the text and of the person judging it~\cite{scheuerman2021severity,zhou2025harm}. The sentiment result needs frame, sentiment, stance, and community together; the crossing result needs a thirteen-dimensional author fingerprint; and the contestation result is the failure of the one-layer content score to read. In this composition, community organizes what is contested, and frame organizes the affect a comment carries, and a method that reduces discourse to one text score misses both.

However, one pair of results looks contradictory. The toxicity label barely moves contestation, yet replies that answer their parent in another frame score 0.0284 higher on Perspective's toxicity score than replies that keep the parent's frame. Both hold, because they describe different things: toxicity accompanies one move inside a conversation, the answering of one frame with another, while contestation is associated with the pairing of frame and community, which describes how a comment is received. What we expect to carry over to other conflicts is the ordering of these factors, not the particular combinations at the top of it. Without Operation Sindoor, our largest event, the three alignment groups keep their order, and humanitarian framing stays well above economic framing. Even within this conflict, the most contested frame differs between alignment groups. A study of another conflict should therefore expect to re-estimate which combinations are contested, and can test whether community and frame again outrank toxicity.

\subsection{Humanitarian framing as the seam}

Humanitarian framing is the frame along which these communities come apart. It is the most contested frame overall (10.45\%), and under it 14.3\% of India-aligned comments read as positive against 5.1\% in general-audience communities, the widest India-aligned gap of any frame. Within communities, it splits two alignment groups and not the third. India-aligned communities contest humanitarian comments at 7.2\%, against 3.9\% for security and military comments, and general-audience communities at 17.4\% against 9.7\%, while \texttt{r/pakistan} contests the two frames at nearly the same rate (8.6\% and 8.5\%) and religious and cultural framing the most (9.3\%). We speculate that security and military framing invites a rally-round-the-flag response that an aligned community can give in unison, while humanitarian framing asks for the moral evaluation that \citet{entman1993framing} counts among a frame's functions, which members of one side can answer differently. The order could also run the other way, with members who already dissent from their community choosing humanitarian framing to say so, as when a flagged India-aligned comment accused India's army of crimes against Kashmiris. Either reading places humanitarian framing where an India-aligned community's agreement breaks, and the \texttt{r/pakistan} exception is consistent with the expectation that a frame's effect depends on what its audience already believes~\cite{chongdruckman2007framing}.

The competing reading is that general-audience communities look contested because their audience is transient, so that occasional participants split the vote mechanically. The reading fails for the commenters we observe. The archive does not record voters. Anyone who reads a comment can vote on it, members or not, and outsiders mobilized from another community can flood a thread~\cite {kumar2018conflict}, while votes themselves are scarce~\cite{gilbert2013underprovision}, so a general-audience vote split may still reflect a mixed crowd of readers rather than disagreement among the community's members. Separating a mixed crowd from a divided membership needs votes broken down by whether each voter belongs to the community, a record that only Reddit holds and could share with researchers in aggregate.

\subsection{Who crosses, and what it costs}

We set out to find bridge users and found none. In our corpus, crossing is a distributed behavior, much as activity across communities runs from generalists to specialists along one continuum~\cite{waller2019generalists}, and the crossers came disproportionately from its combative end. They scored lower on our bridging index than authors who stayed in one community, and the same authors raised their anti-out-group stance when they posted across the boundary. They also lowered their targeted toxicity slightly, though that change rests on a classifier that separates targeted from untargeted toxicity only moderately well. These results bear on two questions that earlier work on crossing left open. The first is what a receiving community penalizes. \citet{datta2019interconflict} inferred conflict from authors rewarded in one subreddit and penalized in another, but could not trace a penalty to what was said. In our data, the penalty did not single out hostility, since outsiders who provoked and outsiders who merely asked questions were both scored down. On \texttt{r/politics}, cross-cutting replies that expressed warmth tended to draw positive responses~\cite{marchal2022benice}; in a community aligned against the poster during a shooting war, even a question was penalized. The second question is whether crossing changes how people write or only who crosses, and we see both. Hawkish authors crossed more often, as a spiral of silence predicts if only a committed core keeps speaking in a hostile climate~\cite{noelleneumann1974spiral}, and the same authors wrote more oppositionally once across, which fits accounts of antisocial behavior as partly situational~\cite{cheng2017troll}.

Contact theory expects contact to reduce prejudice most when groups meet as equals, share goals, cooperate, and have institutional support~\cite{pettigrew2006contact}. Posting into a community aligned against one's own side during a shooting war meets none of these conditions, and what we observe resembles negative contact, which weighs more than positive contact~\cite{paolini2010negative,barlow2012contact}. Twitter users who bridged echo chambers paid a price in network centrality and in how their content was received~\cite{garimella2018echo}; in our data the price fell on outsiders whether or not they bridged. Whether cross-party contact on Reddit integrates or segregates users depends on its conditions~\cite{xia2025integrated}, and ours are conditions under which it went badly. We argue that a system that ranks, recommends, or badges cross-community participation as a proxy for constructive participation would therefore promote incursions, and for this reason we call these participants boundary-crossers rather than bridge users. Whether boundary-crossers give way to contact once the fighting stops can be tested on these authors, because our corpus also holds the comments they post outside the six crisis windows. 

\subsection{Design implications}
\label{sec:disc-design}

Our findings suggest several considerations for the people who design and moderate these communities. First, a moderation queue ranked by toxicity score ranks by the layer that moved contestation least in our data. A queue keyed to frame and community would instead surface the places where reception divides, such as humanitarian framing in general-audience communities or religious framing in \texttt{r/pakistan} during a crisis, and because contestation behaved as a property of the alignment group, such a model could supply the middle level missing from Crossmod, which has classifiers for single subreddits and for the whole site but none for groups of subreddits~\cite{chandrasekharan2019crossmod}. However, a contested comment is not necessarily a harmful one. Most flagged humanitarian comments in general-audience communities carried no toxicity label and disputed factual claims, so a queue that fed removal would suppress legitimate argument, and its cells would need re-estimating for any other conflict. Second, an interface could show, beside a thread, how its frames are voiced in other communities, so readers who take a strike as a source of pride see that general-audience readers attached their positive sentiment to de-escalation. Making that asymmetry visible may harden identity rather than soften it, as exposure to opposing views did for Republican participants in a field experiment on Twitter~\cite{bail2018exposure}, so such a design would need testing before anyone deployed it. Third, because receiving communities penalized outsiders who asked questions as well as those who provoked, a guaranteed period of visibility for out-group comments would protect questions that early downvotes would otherwise bury. The same floor would also protect provocateurs, who are over-represented among crossers. Conditioning it on the register signals our author-fingerprint measures might separate the two. Finally, an event's mix of frames and affect tracked the kind of crisis more closely than the year, so a community could prepare its settings by crisis type, expecting security framing to dominate a military exchange and political framing a constitutional change. Our data cannot identify causal effects of crisis type (Appendix~\ref{app:causal}), the regularity held only for the contrast between military and constitutional events, and two of our three crisis types rest on one event each, so we offer this as a hypothesis for a platform-side experiment rather than as a recommendation.

\subsection{Few clusters and the limits of crisis-response causal inference}

Neither of our difference-in-differences strategies yielded credible identification, and the underlying problems are likely to affect other studies of crisis response. Of their 45 estimates, 15 are significant under cluster-robust (CRV1) inference and none survives a wild cluster bootstrap (Figure~\ref{fig:crv1-wcb}). Because treatment varies by community, these designs have seven to nine clusters, a regime in which cluster-robust standard errors over-reject~\cite{cameron2008bootstrap} and the wild cluster bootstrap itself can be unreliable when few clusters are treated~\cite{mackinnon2017wildbootstrap}. Identification fails as well. Balakot came twelve days after Pulwama, and Operation Sindoor about two weeks after Pahalgam, so each retaliation's baseline is an attack's aftermath and the parallel-trends tests reject. This event clustering is a confounder distinct from the staggered timing that recent difference-in-differences methods address~\cite{goodmanbacon2021did}, and a conflict that escalates in attack-and-retaliation pairs produces it whenever the retaliation falls inside the attack's window.

The spiral of silence, which fits who crosses, is the theory our withdrawal design was built to test, and it stays open in both directions. In each of the four events with a usable author panel, minority-alignment authors cut their posting in the acute phase more than majority authors do, as the theory predicts, minority-alignment comments in that phase are less often neutral than majority comments, which runs against it. Neither pattern survives the bootstrap. We argue that researchers who study crisis response on platforms should report a wild cluster bootstrap beside CRV1 when treatment varies across as few communities as ours, and should check whether their events fall inside one another's windows before trusting a parallel-trends test. We could test whether a hostile opinion climate silences minority commenters by defining treatment at the thread level, where our co-occurrence analysis alone spans 4,210 threads. Our released labels carry each comment's parent and post, so such a design can be built on these six crises without new annotation.

\section{Limitations}

We chose each labeler and checked its prompts on the same gold samples on which we report its agreement, so Table~\ref{tab:reliability} likely overstates its accuracy. Our coders and prompts also saw each item's community and crisis event, so that labels may reflect expectations about them. We did not correct our estimates for labeling error~\cite{angelopoulos2023ppi,egami2023surrogates}, so small group differences in sentiment, stance, or toxicity, where our coders agreed only moderately, may be labeling error. We scored every comment with Perspective set to English, although toxicity classifiers can rate some varieties of English as more toxic~\cite{sap2019racialbias}. Dropping the 10,331 Perspective-scored comments that contain Hindi or Urdu narrows the India-aligned lead over \texttt{r/pakistan} in the share of toxic comments from 2.1 to 1.7 points. The \texttt{controversiality} flag counts a comment that nearly every voter rejected, and one with too few votes, as uncontested, so low contestation does not show agreement. Our ten communities hold only part of Reddit's discussion of these crises, and none centers on Kashmir, although four of our six events concern it. Operation Sindoor supplies 43\% of our analytic sample, nearly thirteen times Pulwama's count, and we cannot separate interest in that crisis from community growth and archive coverage. Last, our author-level results cover only the 5,924 authors with at least ten comments, fewer than one in ten of our 68,517 named authors.

\section*{Positionality Statement}

Our team includes members who grew up in India and outside South Asia. None of us posts in or moderates the communities we study. Although we apply one codebook to every community, this positionality may affect some of our interpretations, particularly of \texttt{r/pakistan}. We make our labels available for examination and alternative interpretations.

\section*{Use of Generative AI}
We used generative AI as a research instrument and as an assistant. As an instrument, gpt-5.2 screened submissions for relevance and labeled content type, narrative frame, sentiment, and stance; Appendix~\ref{app:prompts} gives its prompts and settings. As an assistant, and separately from that benchmark, we used Claude Code to write and debug analysis code. We confirm that the authors wrote all the text in this paper. We used an AI-based writing assistant, Grammarly, solely for grammar and spelling checks and to improve clarity.

\bibliographystyle{ACM-Reference-Format}
\bibliography{paper_cscw}

\appendix

\section{Collection queries}
\label{app:queries}

We collected submissions posted between January 2019 and June 2025 by keyword
search within each of the ten communities. Table~\ref{tab:queries} lists the 39
queries, which we group by topic, with the number of distinct submissions each
matched. A submission can match several queries, so the counts sum to more than
the 38,744 submissions they cover. Our collection record lists the matched
queries for 38,744 of the 43,713 submissions we collected; for the other 4,969 it does not say which queries they matched, so every count in the table is a lower bound.

\begin{table}[htbp]
\centering
\caption{Collection queries, grouped by topic, with the number of distinct
submissions each matched.}
\label{tab:queries}
\small
\begin{tabular}{@{}lr@{}}
\toprule
Query & Submissions \\
\midrule
\multicolumn{2}{@{}l}{\textit{India and Pakistan named together}} \\
\quad India Pakistan & 17,436 \\
\quad Modi Pakistan & 1,515 \\
\quad Imran Khan India & 855 \\
\quad India-{}Pakistan & 580 \\
\quad Indo-{}Pak & 382 \\
\quad Indo Pakistan & 322 \\
\quad Bharat Pakistan & 255 \\
\quad Sharif India & 186 \\
\quad bilateral talks & 183 \\
\quad Hindustan Pakistan & 137 \\
\addlinespace
\multicolumn{2}{@{}l}{\textit{Kashmir}} \\
\quad Kashmir & 4,111 \\
\quad Jammu Kashmir & 3,386 \\
\quad Article 370 & 1,902 \\
\quad Kashmir India Pakistan & 1,895 \\
\quad Kashmir conflict & 446 \\
\addlinespace
\multicolumn{2}{@{}l}{\textit{Attacks}} \\
\quad Pulwama & 2,516 \\
\quad Pahalgam & 291 \\
\quad Jaish & 174 \\
\quad CRPF attack & 93 \\
\addlinespace
\multicolumn{2}{@{}l}{\textit{Military action}} \\
\quad Balakot & 1,105 \\
\quad war & 1,091 \\
\quad surgical strike & 710 \\
\quad cross-{}border & 481 \\
\quad IAF Pakistan & 444 \\
\quad Operation Sindoor & 396 \\
\quad airstrike & 237 \\
\quad airstrike India & 152 \\
\quad PAF India & 86 \\
\quad military strike & 83 \\
\addlinespace
\multicolumn{2}{@{}l}{\textit{Border and ceasefire}} \\
\quad LOC & 1,336 \\
\quad Line of Control & 1,056 \\
\quad ceasefire & 1,006 \\
\quad ceasefire violation & 361 \\
\quad DGMO & 6 \\
\addlinespace
\multicolumn{2}{@{}l}{\textit{Nuclear weapons}} \\
\quad nuclear war & 1,826 \\
\quad nuclear India & 1,752 \\
\quad nuclear Pakistan & 1,026 \\
\quad nuclear threat & 812 \\
\quad nuclear & 108 \\
\midrule
Submissions matching any query & 38,744 \\
\bottomrule
\end{tabular}
\end{table}

\section{Codebook}
\label{app:codebook}

Our coders labeled with a written codebook and made the targeting decision with a separate set of rules. Below we condense both to the
layers this paper reports. For each layer we give the unit coded, the labels with their definitions, and the rule that decides between labels. Our model prompts draw on the same definitions and rules (Appendix~\ref{app:prompts}).

\begin{enumerate}
\item \textbf{Relevance} (one label per submission).
\begin{itemize}
\item \textit{Relevant}: the submission meaningfully engages with relations between
India and Pakistan, their conflict events, tensions between them, or directly
related regional issues such as the Kashmir dispute.
\item \textit{Irrelevant}: the submission does not substantively engage with that
relationship, even if it mentions India or Pakistan in passing.
\end{itemize}
The deciding question is whether the relationship between the two countries is
the subject of the submission. A passing mention of either country, a third
country's politics, or a link to the conflict that takes heavy inference to see
makes a submission irrelevant.

\item \textbf{Content type} (one label per comment).
\begin{itemize}
\item \textit{Factual claim}: asserts a proposition about the real world that can
in principle be checked against evidence, and does nothing else. The test is
whether the sentence could appear as written in a news-wire report.
\item \textit{Opinion or commentary}: expresses a belief, evaluation,
interpretation, or moral judgment, including a prediction. A comment that cites
facts in order to argue, evaluate, or persuade is an opinion.
\item \textit{Question}: has an interrogative structure and seeks information,
including rhetorical questions.
\item \textit{Satire or humor}: shows explicit satire markers (such as
\texttt{/s}), meme-like exaggeration, irony, or parody, with no serious intent to
assert a fact.
\end{itemize}
Coders ask three questions in order. Is the comment primarily a question? Then it
is a \textit{question}. Is its primary intent satire, humor, or parody? Then it is
\textit{satire or humor}. Does it assert a verifiable claim with no evaluative or
argumentative framing at all? Then it is a \textit{factual claim}; otherwise it is
\textit{opinion or commentary}. When facts and evaluation occur together, the
comment is an opinion: if stripping the evaluative language removes the
comment's point, it was an opinion. When in doubt between a factual claim and an
opinion, coders choose opinion. In our analysis we collapse this layer to opinion
against the other three labels.

\item \textbf{Narrative frame} (one dominant frame per comment).
\begin{itemize}
\item \textit{Security and military}: the conflict as armed confrontation, military
action, retaliation, deterrence, or a threat to national security. Includes
terrorism, nuclear escalation, and a ceasefire framed as the end of military
action.
\item \textit{Humanitarian}: the conflict through civilian suffering, human-rights
violations, displacement, or responsibility toward non-combatants.
\item \textit{Political}: the conflict through government decisions, parties,
leaders, constitutional change, elections, or regime responsibility, including
diplomacy framed as a political process.
\item \textit{Economic}: the conflict through economic harm or benefit, sanctions,
trade, development, or the cost of war.
\item \textit{Historical grievance}: the conflict as rooted in past injustice, with
history as the main explanation, for example Partition and 1947.
\item \textit{Religious or cultural}: the conflict as a clash of religious
identities or cultural values, with religion or culture as the central
explanation.
\end{itemize}
Coders choose the frame that structures the comment's explanation of the
conflict, not the one mentioned most often or last, by asking why, according to
the comment, the conflict is happening. A comment that mentions military
operations but centers on civilian casualties is \textit{humanitarian}. Explicit
religious identity markers (such as Islam, Muslim, Hindu, jihad, or Hindutva)
used to frame the conflict route to \textit{religious or cultural}, even when
political actors are named. If a comment is too general to frame on its own,
coders use the title of its submission as the anchor. An earlier version of the
codebook had eight frames; after the pilot round we merged nuclear threat into
\textit{security and military} and divided peace and diplomacy between
\textit{political} and \textit{security and military}.

\item \textbf{Sentiment} (one label per comment).
\begin{itemize}
\item \textit{Positive}: support, approval, hope, optimism, relief, or another
favorable evaluation.
\item \textit{Negative}: criticism, disapproval, anger, frustration, pessimism,
concern, or condemnation.
\item \textit{Neutral}: factual reporting, a balanced presentation of several
perspectives, or no clear evaluative position.
\end{itemize}
Coders code sarcasm by its intended meaning, so ``Oh great, another surgical
strike'' is \textit{negative}. When sentiments mix, they assign the dominant one,
and a truly balanced comment is \textit{neutral}.

\item \textbf{Stance toward India} and \textbf{stance toward Pakistan} (one label
per comment on each axis).
\begin{itemize}
\item \textit{Pro}: support, defense, justification, or sympathy toward the
country, its government, its military, or its people, including praise and
positive framing.
\item \textit{Anti}: criticism, condemnation, blame, or hostility toward the
country, its government, its military, or its people, including negative framing
and accusations.
\item \textit{Neutral}: the comment mentions the country without a clear stance,
reports facts, is genuinely balanced, or gives too little to infer a stance.
\end{itemize}
Coders label the two axes independently: supporting India does not imply opposing
Pakistan, and all nine combinations are valid. Stance can be implicit. Praising
``the Indian military's restraint'' is \textit{pro} India even if the comment never
names India, and criticizing ``ISI interference'' is \textit{anti} Pakistan.

\item \textbf{Toxicity check} (one label per comment, for the 200 comments sampled
on both sides of the threshold of
Perspective~\cite{jigsaw2018perspective,lees2022perspective}).
\begin{itemize}
\item \textit{Offensive, targeted}: inappropriate language, insults, or threats
directed at a specific individual or group.
\item \textit{Offensive, untargeted}: profanity, vulgar language, or other
offensive content without a clear target.
\item \textit{Not offensive}: none of the above.
\end{itemize}
In the corpus, the decision between toxic and not toxic is Perspective's threshold
of 0.5. To score that threshold against the coders, we count both offensive labels
as toxic.

\item \textbf{Target} (one label per comment, for comments Perspective scores as
toxic). The question is whether the offensive language attacks an identifiable
person, group, nation, or religion. Coders apply the tests below in order and stop
at the first that holds.
\begin{itemize}
\item \textit{Named entity}: the insult names, or is aimed at, a person, country,
political party, or organization (``Modi is a fascist dictator'', ``ISI funds
terrorism''). Targeted.
\item \textit{Pronoun}: \textit{you} or \textit{your} in the scope of an insult
attacks the person being addressed, and \textit{they} or \textit{them} referring
to a group in the scope of an insult attacks that group. Targeted.
\item \textit{Group noun}: a demonym, a religious, caste, or ethnic term, or a
political label (such as \textit{bhakt} or \textit{sanghi}) in the scope of an
insult. Targeted.
\item \textit{Coded slur}: a coded slur, deliberate misspelling, or dehumanizing
metaphor (vermin, disease) for an identifiable group, even when the tone is
informal or joking. Targeted.
\item \textit{Sarcasm}: sarcasm or irony that attacks an identifiable group despite
positive surface words. Targeted.
\item \textit{Intensifier}: profanity that intensifies or vents about a situation,
an event, or an abstract idea rather than an entity. Untargeted. Profanity aimed
at a group or an institution is targeted.
\item \textit{Unclear reference}: if \textit{they}, \textit{them}, or \textit{these
people} can be resolved to an identifiable group from the submission title or
earlier sentences, targeted; otherwise untargeted.
\end{itemize}
A comment that quotes someone else's offensive speech is labeled by the
commenter's own position: quoting it to criticize it is untargeted, and endorsing
the quoted attack is targeted. When the target cannot be determined, coders
choose untargeted.
\end{enumerate}

\section{Labeler benchmark}
\label{app:benchmark}

Table~\ref{tab:benchmark} scores each system we tested against the adjudicated
reference for its layer and lists our two coders' agreement beside it for
comparison. It uses the full annotation samples, which we drew before we fixed the ten communities, so its coder values differ slightly from Table~\ref{tab:reliability}; because the samples differ by layer, we compare
systems only within a layer. The table leaves out systems that failed rather than underperformed, such as the three stance classifiers built on natural-language inference, which put every comment in one class.

For content type and narrative frame, we tested three sizes of OpenAI's GPT-5 family (gpt-5.2, gpt-5-mini and gpt-5-nano), each prompted zero-shot and with chain-of-thought reasoning~\cite{wei2022chain}, and each run through both the Chat Completions and the Responses API, twelve conditions per layer, on the 250-item adjudicated gold. gpt-5.2 zero-shot scored highest on both layers ($\kappa = 0.599$ on content type and 0.688 on narrative frame). For sentiment and stance we tested a lexicon baseline VADER~\cite{hutto2014vader}), pre-trained sentiment encoders (among them XLM-T~\cite{barbieri2022xlmt} and BERTweet~\cite{nguyen2020bertweet}), zero-shot stance classifiers built on natural-language inference~\cite{yin2019benchmarking} with BART~\cite{lewis2020bart} and DeBERTa-v3~\cite{he2023debertav3,laurer2024less} encoders, and open-weight instruction-tuned models (among them Gemma-3-12B, Llama-3.1-8B, Mistral-7B-v0.3 and Phi-4-mini) at 4-bit precision under five prompting strategies: zero-shot, few-shot~\cite{brown2020language}, chain-of-thought, self-consistency~\cite{wang2023selfconsistency} and a two-step prompt. The best open-weight result on sentiment, Mistral-7B-v0.3 with self-consistency ($\kappa = 0.526$), rose to 0.566 with parameter-efficient tuning (QLoRA~\cite{dettmers2023qlora}), close to the coder pair's 0.549. On stance, no open-weight system came close (0.407 toward India and 0.455 toward Pakistan, against 0.521 and 0.557 for the coders). We also ran gpt-5.2, gemini-3.1-pro-preview, and claude-opus-4.6 zero-shot on sentiment and both stance axes. All three achieved $\kappa$ between 0.66 and 0.78, above every open-weight system, and on each layer they differed by at most 0.10; GPT-5.2 reached 0.721 on sentiment and 0.664 and 0.775 on stance toward India and Pakistan. These three models also voted on the disputed sentiment and stance items in this reference, favoring them over the open-weight systems. For target type, we tested thirteen zero-shot systems, six fine-tuned encoders under five-fold cross-validation, and two parameter-efficient fine-tunes. A fine-tuned RoBERTa-large performed the best ($\kappa = 0.487$), but it was still below the coders' 0.682. Gemma-3-12B tuned with QLoRA came close (0.480), but we could score it only on a single held-out split of 100 items. Perspective, scored on the same decision, was the worst of the systems that completed it (0.228).

\section{Labeling prompts}
\label{app:prompts}

We labeled relevance and the four comment layers with
gpt-5.2~\cite{openai2025gpt52} through OpenAI's Batch
API~\cite{openai2026batchapi}, with reasoning effort set to none, temperature 0,
and a strict JSON schema. Each call returned one label from the layer's
categories, a confidence between 0 and 1, and a one-sentence reason, and the
stance call returned a label and a confidence for each country. We labeled
content type, narrative frame, and sentiment in separate calls and both stance
axes in one call. Every prompt is zero-shot, in that it contains no comments from
our corpus. Each system prompt, of roughly 700 to 800 words (about 300 for the
relevance gate), states the task, defines every label as the codebook does
(Appendix~\ref{app:codebook}), gives the codebook's decision rule, and lists
boundary cases illustrated with short invented phrases, some of them paired with
the label they should receive. It ends by fixing the output format. The user
prompt then supplies the item to label and its context. Table~\ref{tab:prompts}
summarizes what each call received and the rules its system prompt stresses. 

Our production prompts are longer than the zero-shot prompts we benchmarked: they add boundary rules from the final version of our codebook, among them, for content type, a narrower definition of a factual claim. In each prompt, we gave the model the comment, its post title, and body; in the content-type and frame prompts, we also gave the subreddit and crisis-event context our coders saw (Appendix~\ref{app:prompts}). We re-scored the revised prompts on the same gold before the full run and kept them. Content type enters our analysis only in its collapsed form, as opinion against everything else.

\begin{table*}[htbp]
\centering
\caption{What each labeling call received and the rules its system prompt
stresses.}
\label{tab:prompts}
\footnotesize
\begin{tabularx}{\linewidth}{@{}p{2.3cm}p{5.2cm}X@{}}
\toprule
Call & Context in the user prompt & Rules stressed in the system prompt \\
\midrule
Relevance (per post) & Subreddit; crisis type and phase; score and comment
count; title; first 100 words of the body & Relevance means the post engages
with relations between the two countries; score and comment count may decide only
borderline cases \\
\addlinespace
Content type & Subreddit; crisis event; post title; first 300 words of the
post body; the comment & Three ordered questions (question, satire, factual
claim); a comment that mixes facts with evaluation is an opinion \\
\addlinespace
Narrative frame & As for content type & The frame that structures the comment's
explanation of the conflict; religion as explanation against religion as
descriptor; the post title as anchor for short comments; routing of nuclear
and peace references \\
\addlinespace
Sentiment & First 200 characters of the post title; first 300 characters of
its body; first 1,500 characters of the comment & Sarcasm by intended meaning; the
dominant sentiment when sentiments mix; even mild evaluation is not neutral \\
\addlinespace
Stance (both countries) & As for sentiment & The two axes are labeled
independently; implicit cues count; criticism of a government counts as stance
against its country \\
\bottomrule
\end{tabularx}
\end{table*}

\section{Event signatures under three distance metrics}
\label{app:signature}

An event's signature is the joint distribution of its comments over six frames
and three sentiments, pooled across communities. We compared signatures by cosine
distance, by the Jensen--Shannon divergence, and by Euclidean distance, each with
95\% intervals from the same bootstrap over threads (1,000 resamples).
Table~\ref{tab:signature-metrics} gives every pair of events under all three. Its
last row gives the gap between the two families of pairs: the smallest lower
bound among pairs of a military event and the constitutional change, minus the
largest upper bound among military pairs. Under each metric, every pair of
military events sits closer than every pair of a military event and the
constitutional change, and the gap is positive, so the intervals of the two
families do not overlap. The 2021 ceasefire sits as close to the constitutional
change, under every metric, as the military events sit to one another, and its
pairs with military events fall between the two families.

\begin{table*}[htbp]
\centering
\caption{Distances between event signatures under three metrics, with 95\%
intervals from a bootstrap over threads (1,000 resamples). The gap is the
smallest lower bound among military and constitutional pairs minus the largest
upper bound among military pairs.}
\label{tab:signature-metrics}
\scriptsize
\setlength{\tabcolsep}{3.5pt}
\begin{tabular}{@{}lrrr@{}}
\toprule
Event pair & Cosine distance & Jensen--Shannon divergence & Euclidean distance \\
\midrule
\multicolumn{4}{@{}l}{\itshape Military pairs} \\
\quad Sindoor--Pulwama & 0.0074 [0.0022, 0.0313] & 0.0064 [0.0040, 0.0195] & 0.0549 [0.0298, 0.1116] \\
\quad Pahalgam--Pulwama & 0.0125 [0.0071, 0.0372] & 0.0187 [0.0127, 0.0359] & 0.0673 [0.0520, 0.1161] \\
\quad Sindoor--Pahalgam & 0.0162 [0.0097, 0.0354] & 0.0232 [0.0158, 0.0366] & 0.0803 [0.0623, 0.1170] \\
\quad Balakot--Sindoor & 0.0247 [0.0159, 0.0378] & 0.0103 [0.0074, 0.0163] & 0.0979 [0.0782, 0.1216] \\
\quad Balakot--Pulwama & 0.0260 [0.0159, 0.0468] & 0.0166 [0.0113, 0.0308] & 0.0971 [0.0754, 0.1306] \\
\quad Balakot--Pahalgam & 0.0492 [0.0322, 0.0768] & 0.0361 [0.0256, 0.0547] & 0.1325 [0.1068, 0.1684] \\
\addlinespace
\multicolumn{4}{@{}l}{\itshape Military and constitutional pairs} \\
\quad Article 370--Pulwama & 0.2058 [0.1359, 0.2668] & 0.0939 [0.0673, 0.1240] & 0.2764 [0.2217, 0.3206] \\
\quad Article 370--Balakot & 0.2131 [0.1606, 0.2685] & 0.1085 [0.0817, 0.1383] & 0.2799 [0.2413, 0.3180] \\
\quad Article 370--Pahalgam & 0.2487 [0.1907, 0.3060] & 0.1097 [0.0874, 0.1368] & 0.3020 [0.2625, 0.3402] \\
\quad Article 370--Sindoor & 0.2749 [0.2332, 0.3114] & 0.1287 [0.1077, 0.1497] & 0.3248 [0.2958, 0.3493] \\
\addlinespace
\multicolumn{4}{@{}l}{\itshape Pairs with the 2021 ceasefire} \\
\quad Article 370--Ceasefire 2021 & 0.0377 [0.0229, 0.0833] & 0.0295 [0.0215, 0.0555] & 0.1204 [0.0938, 0.1743] \\
\quad Ceasefire 2021--Pulwama & 0.0924 [0.0382, 0.1513] & 0.0536 [0.0272, 0.0856] & 0.1796 [0.1159, 0.2344] \\
\quad Balakot--Ceasefire 2021 & 0.1087 [0.0488, 0.1691] & 0.0659 [0.0357, 0.1015] & 0.1925 [0.1294, 0.2426] \\
\quad Ceasefire 2021--Pahalgam & 0.1245 [0.0674, 0.1917] & 0.0672 [0.0416, 0.1024] & 0.2052 [0.1529, 0.2566] \\
\quad Ceasefire 2021--Sindoor & 0.1414 [0.0812, 0.2001] & 0.0764 [0.0507, 0.1076] & 0.2269 [0.1754, 0.2701] \\
\midrule
Gap between the two families & 0.0591 & 0.0127 & 0.0533 \\
\bottomrule
\end{tabular}
\end{table*}

\section{Frame co-occurrence within threads}
\label{app:cooccurrence}

\begin{figure*}[h!]
  \centering
  \includegraphics[width=.85\linewidth,alt={Frame co-occurrence within threads for each crisis event.}]{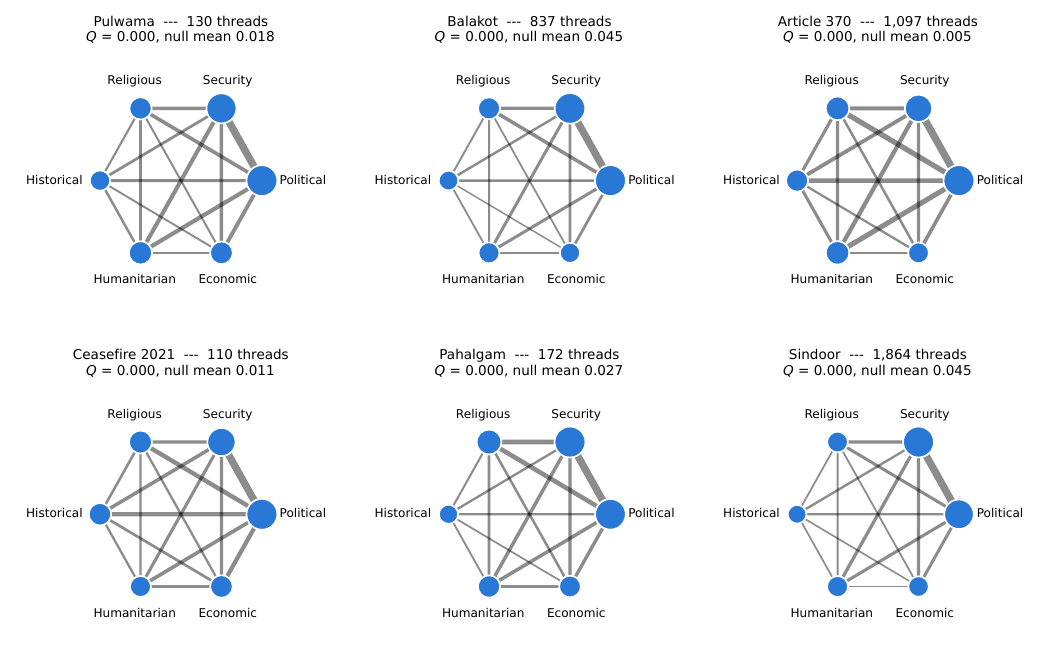}
  \caption{Frame co-occurrence within threads for each crisis event. Node size
  is the number of threads containing the frame, and edge width the number of
  threads containing both frames. Each panel gives the number of threads, the
  Louvain modularity $Q$, and the mean modularity of 500 graphs with shuffled edge
  weights.}
  \Description{Six network diagrams, one per crisis event, each with the same six
  frame nodes arranged in a circle and all fifteen possible edges drawn. Political
  and security nodes are the largest, and the edge between them is the thickest
  in every event. Each panel reports a modularity of 0.000, against null means
  between 0.005 for Article 370 and 0.045 for Balakot and Operation Sindoor.}
  \label{fig:app-cooccurrence}
\end{figure*}

For each event, we linked two frames when both appeared among the comments of one
thread with at least three comments, and weighted the link by the number of such
threads. Figure~\ref{fig:app-cooccurrence} shows the six graphs, which span 4,210
threads. Every pair of frames is linked in every event, so each graph is
complete. Louvain modularity, averaged over 20 seeds, is 0.000 in every event,
against null means between 0.005 and 0.045 for graphs whose edge weights we
shuffled 500 times.

\section{Difference-in-differences designs}
\label{app:causal}

\begin{figure}[!h]
  \centering
  \includegraphics[width=0.6\linewidth,alt={Difference-in-difference design (CRV1).}]{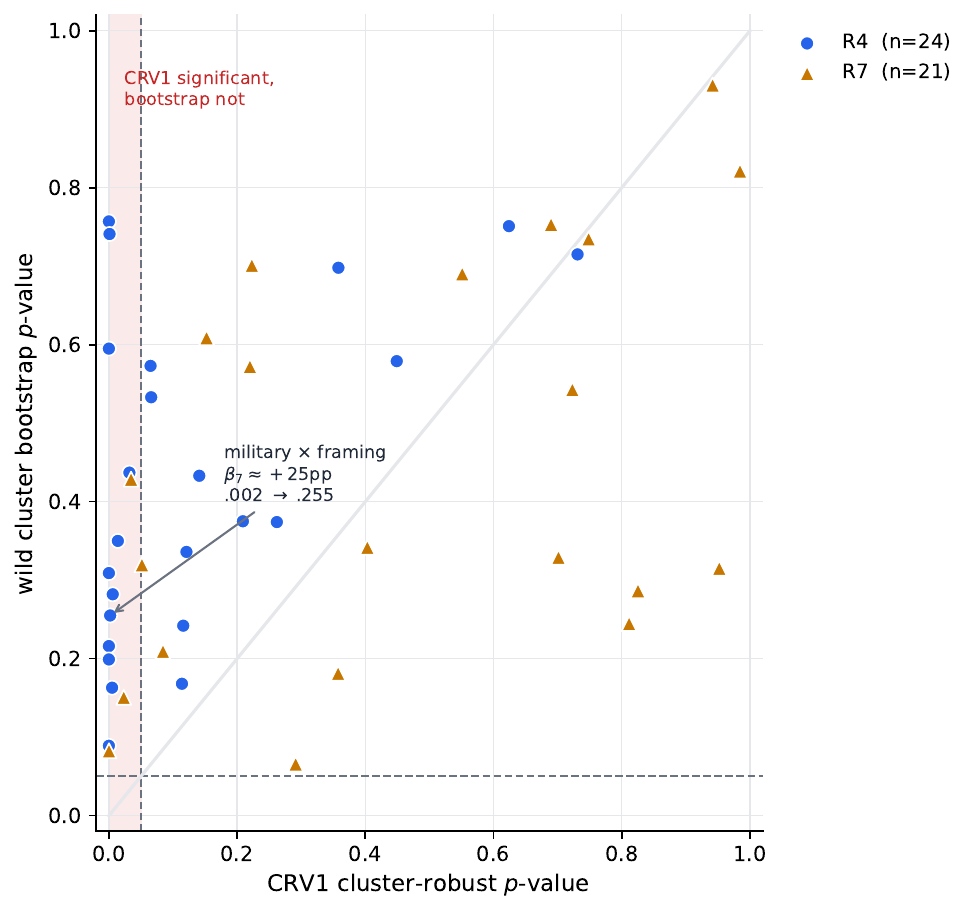}
  \caption{Cluster-robust (CRV1) against wild-cluster-bootstrap $p$-values for the
  45 estimates of the two difference-in-differences designs: 24 from the
  polarization design (R4 in the legend) and 21 from the withdrawal design (R7).
  The shaded column marks CRV1 $p < 0.05$, the dashed horizontal line a bootstrap
  $p$ of 0.05, and the diagonal equal $p$-values. The arrow marks the polarization
  design's crisis-type estimate for security and military framing.}
  \Description{Scatterplot of 45 difference-in-differences estimates, with the
  cluster-robust p-value on the horizontal axis and the wild-cluster-bootstrap
  p-value on the vertical axis. Circles mark the 24 polarization-design estimates
  and triangles the 21 withdrawal-design estimates. A shaded column marks the 15
  estimates with a cluster-robust p-value below 0.05. All 15 lie above the dashed
  line at a bootstrap p-value of 0.05, the lowest near 0.08. An arrow labels the
  estimate for security and military framing after military crises, which moves
  from 0.002 to 0.255.}
  \label{fig:crv1-wcb}
\end{figure}

We ran two difference-in-differences designs, and neither identifies an effect.
The polarization design asked whether a crisis shifts negative sentiment,
security and military framing, and toxicity more in India-aligned communities
than in general-audience ones. The withdrawal design asked whether, in a crisis's
acute phase, comments whose stance labels oppose their community's alignment
(minority-alignment comments) change in length, neutrality, and question rate
more than other comments do, with subreddit and author fixed effects. Both
designs cluster on seven to nine communities, too few for cluster-robust (CRV1)
standard errors alone, so we also computed a wild cluster
bootstrap~\cite{cameron2008bootstrap,mackinnon2017wildbootstrap}. Of the 45
estimates, 15 are significant under CRV1 and none under the bootstrap
(Figure~\ref{fig:crv1-wcb}).

Both designs also fail the parallel-trends test. Of the events we tested in the
polarization design, it rejects parallel trends on all three outcomes for Balakot
and Operation Sindoor, whose baseline phases fall in the aftermath of Pulwama and
Pahalgam, but not for Article 370, and the 2021 ceasefire has too few baseline
comments. In the withdrawal design, it rejects them on all three outcomes (Wald
$\chi^2(2)$ of 48.8, 31.3, and 21.7, all $p < 0.001$). None of the 39 estimates
whose power we computed reaches its minimum detectable effect at 80\% power. In the acute phase, minority-alignment comments become slightly shorter, less
often neutral, and less often questions than majority comments, and only the
drop in neutral comments is significant under CRV1 ($p = 0.023$, bootstrap
$p = 0.15$). In an author panel, minority-alignment authors cut their posting
from the baseline to the acute phase more than majority authors in all four
events with a usable panel (pooled coefficient $-0.143$ on log comment count,
CRV1 $p = 0.005$, bootstrap $p = 0.18$). We therefore make no causal claim.

\end{document}